\documentclass[12pt]{article}
\usepackage[utf8]{inputenc}
\usepackage[margin=3cm]{geometry}
\usepackage{amsmath, amsfonts, amssymb}
\usepackage{color}
\usepackage[breaklinks,colorlinks,urlcolor=blue,citecolor=blue,linkcolor=blue]{hyperref}
\usepackage{graphicx, epstopdf}
\usepackage[T1]{fontenc}
\usepackage[absolute]{textpos}
\usepackage{cite}
\usepackage{tabularx}
\usepackage{subcaption}
\usepackage{appendix}
\usepackage{hyperref}
\usepackage{ragged2e}
\usepackage[capitalise]{cleveref}
\usepackage[labelfont=bf, labelsep=period, font=footnotesize]{caption}
\usepackage{contour}
\usepackage{graphicx}
\usepackage{caption}
\usepackage{subcaption}
\usepackage{float}    

\graphicspath{{FIGS/}}

\newcommand{\wallgo}{\texttt{WallGo~}}

\definecolor{myred}{cmyk}{0,1,1,0.55}
\definecolor{mygreen}{rgb}{0.27, 0.64, 0.48}
\definecolor{mygray}{gray}{.95}

\begin{document}

\vspace{10mm}

\begin{center}
{\bf\Large Benchmarking wall velocities in cosmological\\ phase transitions: Fluid Ansatz and \wallgo} \\
[5mm]
\renewcommand*{\thefootnote}{\fnsymbol{footnote}}
Gl\'auber C. Dorsch$^{a}$\footnote{\href{mailto:glauber@fisica.ufmg.br}{glauber@fisica.ufmg.br}}
Marek Lewicki$^{b,c}$
\footnote{\href{mailto:Marek.Lewicki@fuw.edu.pl}{Marek.Lewicki@fuw.edu.pl}},
and Daniel A. Pinto$^{b}$\footnote{\href{mailto:d.pinto@uw.edu.pl}{d.pinto@uw.edu.pl}}
\\
\vspace{2mm}
$^{a}$\,{\it Universidade Federal de Minas Gerais, 31270-901, Belo Horizonte, MG, Brazil}
\\
\vspace{1mm}
$^{b}$\,{\it Faculty of Physics, University of Warsaw ul. Pasteura 5, 02-093 Warsaw, Poland}
$^{c}$\,{\it Astrocent, Nicolaus Copernicus Astronomical Center Polish Academy of Sciences, ul. Rektorska 4, 00-614, Warsaw, Poland}
\date{}
\end{center}

%%%%%%%%%%%%%%%%ABSTRACT%%%%%%%%%%%%%%%%%%%

\begin{center}
\vspace{1cm}
    {\bf Abstract}
    
\vspace{5 mm}
\justifying{
A reliable computation of the bubble wall velocity during a cosmological phase transition requires an adequate modeling of the non-equilibrium dynamics in the vicinity of this expanding bubble. This task can be made computationally faster by imposing an \emph{Ansatz} on the shape of the non-equilibrium particle distribution function, thus simplifying the collision terms and making the Boltzmann equation solvable in terms of some out-of-equilibrium fluctuations. Two different \emph{Ans\"atze} have prevailed in the recent literature: the so-called fluid \emph{Ansatz} and an expansion in a basis of Chebyshev polynomials, consolidated in the public code \texttt{WallGo}. 
In this work we show that the two approaches yield essentially the same wall velocity in the regime of reasonably mild phase transitions, $\alpha \lesssim 0.01$. Interestingly, the agreement is excellent when only top-quark annihilation is considered, but a noticeable discrepancy appears once scattering processes are included. We also investigate the limitations of linearizing the Boltzmann equation when the fluid \emph{Ansatz} is applied to stronger phase transitions, showing that non-linear contributions induce significant shifts in the predicted terminal velocity as $\alpha\to 1$, even though the non-linear contribution to the wall pressure remain quantitatively small compared to the equilibrium and linearized non-equilibrium parts. We discuss possible consequences of this result for both \emph{Ans\"atze}, while also highlighting the possible limitations of the WKB approach itself when applied to the regime of strong transitions, where the results of both \emph{Ans\"atze} diverge the most.
Since strong phase transitions are precisely the primary targets for future gravitational waves observatories, our study emphasizes that not only higher precision computations of $v_w$ in the semi-classical approach are required, but a treatment beyond the WKB approximation may be needed.}
\end{center}

\renewcommand*{\thefootnote}{\arabic{footnote}}
\setcounter{footnote}{0}

\newpage
\section{Introduction}

We are at the dawn of the gravitational wave (GW) age. It started with the successful detections from ground-based interferometers~\cite{LIGOScientific:2016aoc,
LIGOScientific:2018mvr, LIGOScientific:2021usb, LIGOScientific:2021djp}, followed by announced detections from pulsar-timing arrays~\cite{NANOGrav:2023gor}, and reinforced by the planned launching of the future space-based interferometer LISA for the next decade~\cite{Heffernan:2026kva,LISA:2017pwj}. Naturally, the dawning of this new age urges theoretical advances to parallel the astounding experimental progress. Theoretical predictions must become increasingly more accurate, so as to extract more discerning information from a detected signal and hence have enough resolution to select out a few among the plethora of underlying models that could possibly cause it. 
One possible source of a stochastic GW signal detectable by existing and future observatories is a cosmological first order phase transition, proceeding via the nucleation, expansion and collision of bubbles of a true vacuum in a plasma filled in a metastable state~\cite{Caprini:2018mtu}. Furthermore, aside from this GW spectrum, a number of other relics could also have been produced during the expansion of these bubbles, such as an amount of matter-antimatter 
asymmetry~\cite{Morrissey:2012db, Konstandin:2013caa, Chun:2023ezg} and a dark matter abundance~\cite{Azatov:2021ifm, Baldes:2022oev, Jiang:2023nkj}.

A crucial parameter for accurately determining the observational signatures of all these phenomena is the terminal velocity of the expanding vacuum bubbles. The bubble expansion communicates the phase transition to the surrounding plasma, thus driving the vicinity of the bubble wall out of equilibrium. The particles crossing the phase boundary exert a backreaction force on the scalar field undergoing the phase transition, acting as a hydrodynamical friction that counter-balances the driving pressure of the vacuum energy. Accurately computing this friction requires tracking the non-equilibrium distribution functions of the plasma species, which in the WKB limit is governed by semi-classical Boltzmann equation (applicable when the wall thickness $L_w$ is much larger than the typical particle diffusion length, which essentially amounts to $L_w T\gg 1$ with $T$ the background temperature) ~\cite{Konstandin:2013caa}.

Despite the growing importance of precise predictions for upcoming experiments, solving the full integro-differential Boltzmann equation remains a huge theoretical challenge. Fully numerical solutions do exist~\cite{DeCurtis:2022hlx, DeCurtis:2023hil, DeCurtis:2024hvh,  Branchina:2024rva, Branchina:2025adj} but are time- and resource-consuming, and convoluted techniques must be applied to make the collision terms tractable and computable at a sufficiently fast rate, making this approach extremely challenging and essentially inaccessible to model-builders and phenomenologists. An alternative approach consists in making an \emph{Ansatz} for the distribution functions in terms of some non-equilibrium fluctuations, one that makes the collision terms more tractable, thus reducing the integro-differential equation to a set of ordinary differential equations or even algebraic equations. 

While a number of proposals have been put forward in the literature~\cite{Moore_1995, Cline:2020jre, Kainulainen:2024qpm, Laurent:2020gpg, Laurent:2022jrs, Ekstedt:2024fyq}, two have received particular attention. One is the so-called fluid \emph{Ansatz}~\cite{Moore_1995, Fromme:2006wx, Dorsch:2021, Dorsch:2024jjl, Dorsch:2023tss}, which assumes that the distribution functions has the same form as the Bose-Einstein/Fermi-Dirac distributions, but with a modified functional argument to include parameters such as a chemical potential, temperature and velocity fluctuations, and other perturbations corresponding to dissipative effects. The name ``fluid \emph{Ansatz}'' stems precisely from this possibility of directly mapping the postulated fluctuations onto some hydrodynamical properties of the fluid --- hence one major advantage of this approach. 
Alternatively, another approach has recently appeared in the literature consisting in expanding the distribution function in an orthogonal basis of Chebyshev polynomials~\cite{Laurent:2022jrs}. This approach has been popularized and made widely accessible to the broad community by its implementation in the publicly released code \wallgo\cite{Ekstedt:2024fyq, vandeVis:2025plm}.   

So far no comparison of these two methods for calculating the friction and the bubble wall velocity has ever been made --- it is the purpose of the present work to fill this gap. For the fluid \emph{Ansatz} we compute the collision matrices to leading-log approximation --- neglecting collision processes whose amplitudes are not logarithmically divergent in the infrared~\cite{Arnold:2000dr} ---, but we move beyond some approximations made elsewhere in the literature when computing the integrals~\cite{Dorsch:2021ubz, Dorsch_2022}. The Boltzmann equation then reduces to a coupled system of differential equations which we solve with a custom-made \texttt{Python} code to find the fluctuations, then compute the friction and the wall velocity by solving the equation-of-motion for the scalar field immersed in the plasma. Similarly, \wallgo computes the collision terms also at leading-log approximation using the Chebyshev expansion, and then also computes the wall velocity by solving the Klein-Gordon equation. 
Outcomes of these two methods are compared within two complementary theoretical frameworks: a Standard Model Effective Field Theory (SMEFT) with a $\phi^6$ operator --- which allows for rather strong phase transitions~---, and a Standard Model potential with a fiducial light Higgs $m_h \lesssim 80$~GeV, which also leads to first order phase transitions, albeit weaker than the SMEFT. 

We find that both approaches yield the same results (up to the estimated errors) when the phase transition is sufficiently weak, i.e. when the ratio of energy released by the phase transition to the radiation energy density of the background fluid is $\alpha\lesssim 0.01$. The agreement is remarkable when only the top-quark annihilation process is included. Interestingly, the agreement is noticeably worse when scattering interactions are taken into account. The discrepancy in the wall velocity estimate reaches $\mathcal{O}(40\%-60\%)$ when $\alpha\gtrsim 0.1$. But it is worth emphasizing that in this regime the wall thickness is also $L_w T\lesssim 3$, thus approaching the limit of validity of the WKB approximation. In this regime it is pointless to discuss the viability of either \emph{Ansatz}, since the very modeling of the problem in terms of a Boltzmann equation is questionable. In any case, we further investigate possible causes of this breakdown by studying the impact of incorporating $\mathcal{O}(\text{fluctuations}^2)$ terms in the fluid \emph{Ansatz}, finding that these higher-order corrections induce significant shifts in velocities in a strong phase-transition scenario, even though the $\mathcal{O}(\text{fluctuations}^2)$ contributions to the pressure remains comparatively small relative to the linearized case.

The paper is structured as follows. In Section~\ref{sec:hydrodynamics}, we review the thermodynamical description of cosmological phase transitions and the formalism that determines the friction term acting on the bubble wall. Section~\ref{sec: BEs} introduces the semi-classical Boltzmann equation and details the two distinct approaches utilized to compute the non-equilibrium distribution functions: the extended fluid \emph{Ansatz} and the spectral method based on Chebyshev polynomials implemented in \wallgo. In Section~\ref{sec:models}, we specify the SMEFT and Light Higgs models and present the benchmarking results for the terminal wall velocities, identifying the limits of the linear approximation. Section~\ref{sec:second_order} extends the fluid formalism to incorporate the non-linear (fluctuations)$^2$ contributions, allowing us to investigate the sensitivity of the system to non-linearities during strong phase transitions. Finally, we summarize our main findings and theoretical implications in Section~\ref{sec:conclusions}.

\section{Phase transition and bubble wall friction}
\label{sec:hydrodynamics}
We consider a scenario in which the early Universe undergoes a first-order phase transition via the nucleation and expansion of vacuum bubbles in the primordial plasma. Our goal is to compute the velocity of this bubble expansion --- or, rather, to compare two different approaches towards this computation. As we will see, this will be strongly dependent on the non-equilibrium dynamics. However, many relevant parameters to this computation depend also on equilibrium effects, which we will therefore discuss in this section. 

For simplicity we will consider models with one dynamical scalar field $\phi$ undergoing the phase transition. The explicit models will be discussed in section~\ref{sec:models}. For now it suffices to assume that the free-energy $\mathcal{F}(\phi,T)$ is known as a function of the plasma temperature $T$ and this field $\phi$. The phase transition starts at the temperature $T_n$ when bubble nucleation starts to become efficient compared to the Hubble rate. In order to determine the nucleation temperature, we first compute the Euclidean action $S_3$ associated with bubble nucleation in our potential, and then apply the standard criterion $S_3/T \approx 140$ to identify the transition temperature, see~\cite{Caprini:2019egz} for details. We have also compared this approximation to the more sophisticated treatment of ref.~\cite{Ellis:2018mja} and found only negligible differences for the non-supercooled phase transitions considered in this work. Finally, another important thermodynamical parameter is the ratio of energy released in the transition per radiation energy density in the background fluid, which we quantify as\footnote{For a detailed discussion on different definitions and comparisons among them see also~\cite{Giese:2020rtr,Giese:2020znk}.}
\begin{equation}
    \alpha \equiv \frac{1}{\rho_{\rm rad}} \left(  \Delta \mathcal{F} - \frac{T}{4} \frac{\partial \Delta \mathcal{F}}{\partial T}  \right),
\end{equation}
where $\Delta \mathcal{F}$ is the free-energy difference between the initial (metastable) and the final (stable) vacua. Notice that this parameter $\alpha$ can serve as a quantifier of the strength of the first order phase transition, an intuition which we will use in what follows.

\subsection{Background profile}

As the bubbles nucleate and expand, the plasma around the bubble wall heats up, leading to a spatially-dependent temperature and velocity profiles which can be determined from macroscopic energy-momentum conservation,
\begin{equation}
    \partial^\mu T_{\mu\nu} = \partial^\mu T_{\mu\nu}^{\text{plasma}} + \partial^\mu T_{\mu\nu}^{\phi} = 0.
    \label{eq:momentum_conservation}
\end{equation}
The energy-momentum tensor of the scalar field is
\begin{equation}
    T_{\mu\nu}^{\phi} = \partial_\mu \phi \partial_\nu \phi - g_{\mu\nu} \left( \frac{1}{2}\partial_\sigma \phi \partial^\sigma \phi - V_0(\phi) \right),
\end{equation}
with $V_0$ the zero-temperature vacuum potential. Assuming thermal equilibrium for the background plasma, its energy-momentum tensor reduces to
\begin{equation}
    T_{\mu\nu}^{\text{plasma}} = \omega u_\mu u_\nu - P g_{\mu\nu},
\end{equation}
where $P$ and $\omega \equiv T \partial P / \partial T$ are respectively the pressure and enthalpy of the fluid, and $u_\mu$ is the local four-velocity of the fluid. Notice that once we sum $T_{\mu\nu}^\text{plasma}+T_{\mu\nu}^\phi$ to get the total energy-momentum, the zero temperature potential term $V_0(\phi)$ can be combined with the pressure to yield the total free energy density of the system, $\mathcal{F} \equiv V_0(\phi) - P$.

We will consider a steady-state planar wall, after it reaches a constant terminal velocity $v_w$ along a direction $z$ orthogonal to the wall. Due to self-similarity, the solution depends only on a parameter $\xi = \gamma_\text{bg}(z-v_w t)$. Then, in the rest frame of the bubble wall, the energy-momentum conservation equations~\eqref{eq:momentum_conservation} simplify to $\partial_\xi T^{\xi\xi}= \partial_\xi T^{\xi0} = 0$. By integrating across the $\xi$ direction, we obtain the macroscopic matching conditions for the background with $T_{\rm bg}(\xi), v_{\rm bg}(\xi)$ dependence (cf. ref.~\cite{Dorsch:2024jjl} for details),
\begin{equation}
\begin{split}
    v_{\rm bg}^2\gamma_{\rm bg}^2\omega_{\rm bg} - \mathcal{F}_{\rm bg} + \frac{1}{2}(\partial_\xi\phi)^2 &= k_1, \\
    v_{\rm bg}\gamma_{\rm bg}^2\omega_{\rm bg} &= k_2.
\end{split}
\label{eq:hydro_matching}
\end{equation}
To explicitly find the profile $T_{\rm bg}(\xi)$ and $v_{\rm bg}(\xi)$ across the phase transition coordinate, we parameterize the scalar field with the profile
\begin{equation}
    \phi(\xi) = \frac{\phi_0}{2}\left[1+\tanh\left(\frac{\xi}{L_w}\right) \right].
\end{equation}
This parameterization of the field profile will be used in the rest of the work.  By substituting $\phi(\xi)$ into equation~\eqref{eq:hydro_matching}, the matching conditions become a localized algebraic system. For a given set of boundary conditions $k_1$ and $k_2$, one can solve the system iteratively at each spatial coordinate $\xi$ to yield the continuous background profiles $T_{\rm bg}(\xi)$ and $v_{\rm bg}(\xi)$.

The integration constants $k_1$ and $k_2$ can be determined by evaluating the energy-momentum tensor at the asymptotic boundaries far away from the bubble wall. The specific boundary conditions depend on the hydrodynamical regime of the expanding bubble, which can be uniquely determined by the wall velocity $v_w$. Three cases can be distinguished.

\paragraph{Deflagrations ($v_w<c_s$):}
In these solutions, the expanding wall is preceded by a shock wave that compresses and heats up the plasma. Consequently, the temperature in front of the wall, $T_+$, is strictly higher than the nucleation temperature $T_n$, and the fluid ahead is pushed forward. To properly fix $k_1$ and $k_2$, one must first compare the constant equations~\eqref{eq:hydro_matching} far away from the wall, and then match the velocities and temperatures across the shock front by tracking the jump from $T_n$ to the shock temperature $T_{sh}$, and then evolving the fluid equations of the bubble to find the exact value of $T_+$and $v_+$. The value of $T_+$ and $v_+$ can be plugged back into equation~\eqref{eq:hydro_matching} to find the constants $k_1$ and $k_2$. The detailed procedure of finding $T_+$ and $v_+$ can be found in ref.~\cite{Dorsch:2023tss}.

\paragraph{Hybrids ($c_s<v_w<v_J$):}
Ocurring in the supersonic regime below the Jouguet velocity $v_J$\footnote{This is the minimum wall velocity for a detonation, occurring when the fluid velocity immediately behind the wall, $v_-$, equals the local sound speed in the rest frame of the wall.}, hybrid solutions act as an intermediate stage. They are preceded by a shock wave ahead of the wall, leading to a hydrodynamical heating $T_+>T_n$ similar to the deflagration case, but now with a rarefaction wave trailing behind the bubble as well. For this regime, the boundary conditions are constrained by the requirement that the fluid velocity is equal to the speed of sound ($v_-= c_s$)\footnote{Entropy considerations enforce $v_- \le c_s$, meaning the trailing rarefaction wave must be of the Jouguet type.}. The further procedures to find $T_+$ and $v_+$ follow the same structure as the deflagrations and can be used to find $k_1$ and $k_2$.

\paragraph{Detonations ($v_w>v_J$):}
In this supersonic regime, the wall expands so rapidly that the plasma ahead receives no hydrodynamical information and remains entirely unperturbed. Thus, there is no shock wave, and the boundary conditions are trivially fixed as $T_+=T_n$ and $v_+= v_w$, with a rarefaction wave developing behind the wall to adjust the fluid into the broken phase. In this third trivial case, one can directly plug $T_+$ and $v_+$ in equation~\eqref{eq:hydro_matching} to find the boundary conditions constants $k_1$ and $k_2$ that can be used to find the $T_{\rm bg}(\xi)$ and $v_{\rm bg}(\xi)$.

\subsection{The friction term}
\label{sec:KGeq}

The passage of the bubble drives the plasma away from equilibrium, and this acts on the Higgs field as a counter-pressure against the bubble expansion. Since we want to compute the terminal velocity $v_w$ at which the wall expands, we need to be able to quantitatively account for this pressure. This is usually done by solving the Higgs equation of motion, which can be deduced from the total energy-momentum tensor conservation~\cite{Konstandin_2014, Dorsch_2022}. Separating the equilibrium and out-of-equilibrium contributions of the total energy-momentum conservation leads to~\cite{Konstandin_2014}
\begin{equation}
    - \phi'' + \frac{\partial \mathcal{F}(\phi,T)}{\partial \phi} + \sum_i \frac{d m_i^2}{d \phi} \int \frac{d^3 p}{(2 \pi)^3} \frac{1}{2E_i} \delta f_i (\textbf{p}, x) = 0.
    \label{eq:KG_equation}
\end{equation}
Given our \emph{Ansatz} for the wall profile in equation~\eqref{eq:delta}, the background dynamics is entirely determined by the wall velocity $v_w$ and the thickness $L_w$. Consequently, the system can be constrained by taking two macroscopic moments of eq.~\eqref{eq:KG_equation}, specifically:
\begin{equation}\begin{split}
    M_1 &\equiv \dfrac{1}{T^4}\int_{-\infty}^\infty (\text{LHS of eq.}~\eqref{eq:KG_equation})\,\frac{\partial \phi}{\partial \xi} d\xi =0,\\
    M_2 &\equiv \dfrac{1}{T^5}\int_{-\infty}^{\infty} (\text{LHS of eq.}~\eqref{eq:KG_equation})\,\frac{\partial \phi}{\partial L_w}\,d\xi = 0.
    \label{eq:M1M2_full}
\end{split}\end{equation}
Physically, these two moments enforce the vanishing of the total pressure acting on the wall and of its gradient across the profile, respectively. When the two equations~\eqref{eq:M1M2_full} are satisfied, the bubble maintains a steady-state expansion. Since we employ the profile parametrization without spatial offsets, this system of moments precisely mimics the iterative procedure used in the \wallgo package~\cite{Ekstedt:2024fyq}.

Following ref.~\cite{Konstandin_2014}, these equations can be conveniently recast in a dimensionless form as
\begin{equation}\begin{split}
M_1 &\equiv \dfrac{\Delta V}{T_+^4} +\frac{\mathcal{P}(\delta f)}{T_+^4}  = 0,\\
M_2 &\equiv \dfrac{1}{6 (L_w T_+)^2}\left(\frac{\phi_0}{T_+}\right)^3 + \dfrac{W}{T_+^5} + \frac{\mathcal{G}(\delta f )}{T_+^5} = 0,
\label{eq:M1M2}
\end{split}
\end{equation}
where $\mathcal{P}(\delta f)$ and $\mathcal{G}(\delta f )$ encapsulate the friction terms arising from the integration over the non-equilibrium plasma contributions $(\delta f_i)$. On the other hand the ``driving forces'' are defined as

\begin{equation}
\begin{aligned}
        \Delta V &\equiv \int_{-\infty}^{\infty} d\xi\dfrac{\partial V(T(\xi), \phi)}{\partial \phi}\frac{\partial \phi}{\partial \xi} 
        \\
        W &\equiv  \int_{-\infty}^{\infty} d \xi \dfrac{\partial V(T(\xi), \phi)}{\partial \phi}\frac{\partial \phi}{\partial \xi} \, \phi_0 \xi
\end{aligned}
\label{eq:deltaV}
\end{equation}

Here, $\Delta V$ represents (minus) the pressure difference across the wall due to the free energy released during the phase transition. Thus, the first moment $M_1$ has a straightforward physical interpretation of being the total pressure acting in the bubble that must be exactly balanced by the hydrodynamic friction $\mathcal{P}(\delta f)$. Furthermore, as pointed out in~\cite{Dorsch:2024jjl}, correctly accounting for the varying temperature profile $T(\xi)$ across the wall introduces an additional thermal gradient force, acting as a new source of pressure on the expanding bubble. This is dubbed the ``equilibrium term'' of the friction contribution.

\section{The Boltzmann equation}
\label{sec: BEs}

Apart from this equilibrium contribution to the pressure on the wall, which can be computed entirely from the free-energy $\mathcal{F}(\phi,T)$, there is also a friction term coming from non-equilibrium fluctuations generated by the bubble expansion. Our purpose in this section is to discuss how to account for these fluctuations.

Provided the bubble wall width is much larger than the typical thermal momentum of the incident particles ($L_w \gg 1/p \sim 1/T$), a WKB approximation is applicable. In this regime, it has been shown that the Kadanoff-Baym equations for the two-point function (which encodes information on the particle distribution functions $f_i$) reduce to the semi-classical Boltzmann equation~\cite{Konstandin:2013caa}

\begin{equation}
\mathcal{L}[f_i] \equiv\left( p^\mu \partial_\mu + \frac{1}{2} \partial_\mu m^2 \partial_{p_{\mu}}\right) f_i =- \mathcal{C}[f_i] ,
\label{eq. BEq}
\end{equation}
where $\mathcal{L}$ is the linear Liouville operator and $\mathcal{C}$ is the collision operator, which accounts for interactions among plasma species, given by
\begin{equation}\begin{split}
    \mathcal{C} = \frac{1}{2}\sum_{\text{processes}} \int &\frac{d^3k\, d^3p^\prime d^3k^\prime}{(2\pi)^{9} 2E_k\, 2E_{p^\prime}\, 2E_{k^\prime}} |\mathcal{M}_{pk\to p^\prime k^\prime}|^2\times\\
       & \times(2\pi)^4 \delta^4(p+k-p^\prime-k^\prime)f_{p} f_{k} (1\pm f_{p^\prime}) (1\pm f_{k^\prime}),
\end{split}\end{equation}
with $\mathcal{M}$ the quantum-mechanical amplitude of the collision process (e.g. annihilation or scattering).

The distribution function can be decomposed into a non-equilibrium and an equilibrium contribution,
\begin{equation}
f_i = f_i^{\text{eq}} + \delta f_i,
\label{eq:f_expansion}
\end{equation}
and our goal is to solve the Boltzmann equation for $\delta f_i$. 

Due to the collision term, this integro-differential equation is a huge challenge to be solved even with a fully numerical approach --- convoluted techniques must be applied to reduce the dimensionality of the collision integrals and to make them more tractable and computable at a sufficiently fast rate for practical purposes~\cite{DeCurtis:2022hlx, DeCurtis:2023hil, DeCurtis:2024hvh, Branchina:2025adj}. Due to the absence of public codes implementing all these detailed calculations, this approach remains rather impractical for a phenomenologist that needs the wall velocity not as an end result but as an input for further analyses. An alternative approach to make the collision integrals more manageable --- and also more prone to a physical interpretation, thus enabling a better intuition of the behaviour of the solution --- is to make an \emph{Ansatz} for the out-of-equilibrium terms $\delta f$, expanding it in some basis parametrized by certain fluctuations. There are mainly two competing \emph{Ans\"atze} in the current literature, which we now briefly describe.

\subsection{The Extended Fluid \emph{Ansatz}}
\label{sec:fluidAnsatz}

The fluid \emph{Ansatz} introduces non-equilibrium effects while maintaining the overall shape of the equilibrium distribution functions, modeling the fluctuations as deviations which can be easily mapped to macroscopic fluid properties such as the chemical potential, temperature, fluid velocities and eventually dissipative effects in a non-ideal fluid~\cite{DeGroot:1980dk, Dorsch:2023tss}. More specifically, if $\beta^{-1} \equiv T$ is the temperature of the plasma, $u_\mu$ its four-velocity relative to the observer, and $p^\mu$ the particle four-momentum, then the equilibrium distribution function depends only on $\beta p^\mu u_\mu$ and the perturbation $\delta(x,p)$ is modeled as fluctuations around this quantity as
\begin{equation}
    f(x,p) \equiv \frac{1}{e^{\beta p^\mu u_\mu - \delta(x,p)} \pm 1} \approx f_i^{\text{eq}} + \delta(x,p) (-f_{eq}'),
    \label{eq:fluid_delta}
\end{equation}
where $(-f_{eq}')$ is the derivative of the equilibrium distribution function with respect to $\beta p^\mu u_\mu$. Notice that the second equality involves a linearization approximation which should hold as long as the fluctuations are sufficiently small. We will discuss the validity of this approximation in section~\ref{sec:second_order}. The fluctuations $\delta(x,p)$ can then be expanded in powers of momenta as
\begin{equation}
\delta(x,p) = w^{(0)}(x) + p^\mu w^{(1)}_\mu(x) + p^\mu p^\nu w^{(2)}_{\mu\nu}(x) + \ldots .
\label{eq:delta}
\end{equation}
Expanding the Liouville operator and the collision operator in the Boltzmann equation, and neglecting terms of $\mathcal{O}(w^2)\sim \mathcal{O}(w \partial m^2)$ and higher orders, leads to the linearized Boltzmann equation
\begin{equation}
\begin{split}
&( \partial_\mu w^{(0)} + p^\rho \partial_\mu w^{(1)}_\rho + p^\rho p^\sigma \partial_\mu w^{(2)}_{\rho\sigma} + \cdots )\ p^\mu (-f_{eq}') + \\[1mm]
& \frac{1}{2}\sum_{\text{processes}} \int \frac{d^3k\, d^3p^\prime d^3k^\prime}{(2\pi)^{9} 2E_k\, 2E_{p^\prime}\, 2E_{k^\prime}} |\mathcal{M}_{pk\to p^\prime k^\prime}|^2 (2\pi)^4 \delta^4(p+k-p^\prime-k^\prime)\times\\
&\quad\qquad\qquad\times f_{0p}^{\text{eq}} f_{0k}^{\text{eq}} (1\pm f_{0p^\prime}^{\text{eq}}) (1\pm f_{0k^\prime}^{\text{eq}}) \left(\delta_p + \delta_k - \delta_{p^\prime} - \delta_{k^\prime}\right)= \\
& \left(u^\mu \frac{\partial_\mu m^2}{2T} + p^\mu\frac{p^\rho}{T} \partial_\mu u_\rho - p^\mu p^\rho u_\rho \frac{1}{T^2} \partial_\mu T \right ) (-f_{eq}') .
\end{split}
\label{eq:Boltzmann_linear}
\end{equation}

In order to extract information on the various $w^{(n)}$, one takes moments of the above equation by multiplying it by factors of $p^\alpha p^\beta p^\gamma \cdots$, projecting along the plasma four-velocity $u^\mu$ or along the perpendicular direction $\bar{u}^\mu$, and integrating over the phase space $\int \frac{d^3 p}{(2\pi)^3 E_p}$~\cite{Konstandin_2014, Dorsch:2021, Dorsch_2022, Dorsch:2024jjl}.
Using this procedure, we can reduce the integro-differential Boltzmann equation to a linear system of ordinary differential equations and find the unknown coefficients of the momentum expansion in eq.~\eqref{eq:delta},
\begin{equation}
q = \left(w^{(0)}, T u^\mu w^{(1)}_{\mu}, T \bar{u}^\mu w^{(1)}_{\mu}, T^{2} u^\mu u^\nu w^{(2)}_{\mu\nu}, T^{2} u^\mu \bar{u}^\nu w^{(2)}_{\mu\nu}, T^{2} \bar{u}^\mu \bar{u}^\nu w^{(2)}_{\mu\nu}, \ldots\right)^T,
\label{eq:q_fluc}
\end{equation}
where appropriate factors of temperature $T$ have been absorbed into $q$ to keep the perturbations dimensionless.

Assuming a steady-state planar wall, fluctuations depend exclusively on the planar coordinate $\xi = x^\mu v_\mu$, with $v^\mu = \gamma_w (v_w u^\mu + \bar{u}^\mu)$, $u^\mu u_\mu = 1, u^\mu \bar{u}_\mu=0$, thereby reducing the derivative operator to $\partial_\mu = v_\mu \partial_\xi$.
After taking $D = \frac{(\kappa+1)(\kappa+2)}{2}$ moments of eq.~\eqref{eq:Boltzmann_linear} for a given order $\kappa$ and integrating over the phase space, we arrive at a coupled system of differential equations for the fluctuations,
\begin{equation}
 A\cdot \partial_\xi (q_i) + \dfrac{(L_wT)}{\gamma_w}\Gamma\cdot q_i = S.
 \label{eq:systemBE}
\end{equation}
A closed formula for $S$, $\Gamma$ and $A$ at a given moment order $\kappa$ containing $D = \frac{(\kappa+1)(\kappa+2)}{2}$ fluctuations can be found in Appendix~\ref{app:sources_en_mom}.

To solve the system in eq.~\eqref{eq:systemBE}, one can apply the standard Green's method \cite{Cline:2020jre,Lewicki:2021pgr,Dorsch_2022}, leading to the fluctuations
\begin{equation}
 \begin{aligned}
 q(z) =\, &\chi\cdot \int_{-\infty}^\infty 
				e^{-\lambda  (z-z^\prime)}
				\cdot \theta(\lambda(z-z^\prime))\cdot \text{sign}(\lambda)\cdot \left(\chi^{-1}\cdot A^{-1}\cdot S(z')\right) dz^\prime,
\end{aligned}
\end{equation}
where the coordinate $z = \xi/L_w$ was introduced, and $\chi$ is a matrix whose columns are the eigenvectors of the matrix $A^{-1}\cdot (L_wT)\Gamma/\gamma_w $, with $\lambda$ a diagonal matrix with the corresponding eigenvalues. 

At this point it is worth pausing to point out a major advantage of this fluid \emph{Ansatz}, namely the clear physical interpretation of these fluctuations. Indeed, for an observer at rest with respect to the bubble wall, who sees the plasma particles approaching with velocity $v_w$, the momentum-dependence of the distribution function in eq.~\eqref{eq:fluid_delta} is
\[
    \beta p^\mu u_\mu - \delta = 
    - w^{(0)} + E\left(\dfrac{1}{T} - w^{(1)}_0\right) + p_z (v_w - w^{(1)}_z) + \ldots
\]
so that $w^{(0)}$ behaves as a chemical potential, $w^{(1)}_0$ encodes fluctuations in the background temperature, $w^{(1)}_z$ are local fluctuations in background fluid velocity, while higher-order terms are related to dissipative terms describing a non-ideal fluid~\cite{DeGroot:1980dk}. Likewise, a glance at the explicit form of the collision matrix $\Gamma$ (see e.g. ref.~\cite{Dorsch:2021}) shows that annihilation processes can affect all these fluctuations (and can mix temperature perturbations with chemical potential) while the role of scatterings is essentially to equilibrate temperature and velocity without altering the chemical potential, as expected. We will see this physical intuition in action when discussing the results in section~\ref{sec:results} below.

\subsubsection*{Perturbative control of the linearization procedure}

In most applications found in the literature the fluid \emph{Ansatz} is used together with the linearization shown in eq.~\eqref{eq:fluid_delta}, neglecting terms $\mathcal{O}(\delta^2)$. 
This approximation is valid if 
\[
\delta(x,p) \ll 1.
\]
Assuming the typical momentum scale is $p \sim T$ in equation~\eqref{eq:delta}, the condition that must be satisfied in the dimensionless fluctuations in eq.~\eqref{eq:q_fluc} can be estimated as
\begin{equation}
R(z) = |q_0(z) + q_1(z) + q_2(z) + q_3(z) + \cdots q_{D}(z)| \ll 1
\label{eq:R}
\end{equation}
where $D = \frac{(\kappa+1)(\kappa+2)}{2}$ is the total number of fluctuations for a truncation in momentum expansion at order $\kappa$, retaining all basis elements weighted by $(p^\mu u_\mu)^{a_i} (p^\nu \bar{u}_\nu)^{b_i}$ subject to the constraint $a_i + b_i \le \kappa$ inside the Boltzmann Equations.

Hereafter, the condition $\text{max}|R(z)|\ll1$ will serve as our primary diagnostic parameter to quantify the breakdown of the linearization.

\subsection{\texttt{WallGo}: Expanding in Chebyshev polynomials}
\paragraph{}An alternative approach to the fluid \emph{Ansatz} is to use a spectral method to solve the linearized Boltzmann equation, which constitutes the core framework of the \wallgo package~\cite{Ekstedt:2024fyq}. Instead of expanding the non-equilibrium distribution function in powers of four-momentum, this method projects the momentum dependence of the perturbations into a basis of orthogonal polynomials.

Because of the wall's planar approximation, the spatial and momentum dependence can be fully parametrized by the $\xi$-coordinate (direction of the wall expansion), longitudinal momentum $p_z$, and the magnitude of the transverse momentum $p_k\equiv p_\perp$. To use the Chebyshev polynomials, which are defined strictly on the domain $[-1,1]$, the physical unbounded variables must be mapped to compact coordinates. For the momentum directions, the mappings are
\begin{equation}
    \rho_z(p_z)= \tanh\left(\frac{p_z}{2T}\right), \ \ \ \ \rho_k(p_k) = 1 -2 \exp\left(-\frac{p_k}{T}\right),
\end{equation}
ensuring that the solution vanishes as $|p|\rightarrow \infty$. In the spatial direction, the coordinate distance from the wall $\xi$ maps to $\chi$ using a grid designed to resolve the finite width of the bubble wall and the decay lengths in the tails $(\xi\rightarrow \pm \infty)$.
Therefore, the distribution function for a species $a$ is expanded as 
\begin{equation}
    f^{a}(p,\xi) = f^{a}_{eq} + \delta f ^{a} (\chi,\rho_z,\rho_k) = f^{a}_{eq} + \sum_{i=2}^{M}\sum_{j=2}^{N}\sum_{k=1}^{N-1}\delta f^{a}_{ijk} \bar{T}_i(\chi)\bar{T}_j(\rho_z)\tilde{T}_k(\rho_k),
    \label{eq:f_wallgo}
\end{equation}
where $M$ and $N$ represent the number of basis polynomials in the spatial and momentum directions, respectively, and $\delta f_{ijk}^{a}$ are the coefficients to be determined. To ensure the boundary conditions $\delta f^a \rightarrow 0$ at the infinities of spatial and momentum variables, the method uses restricted Chebyshev polynomials,
\begin{equation}
\bar{T}_i(x) = \begin{cases} T_i(x) - T_0(x), & i \text{ even} \\ T_i(x) - T_1(x), & i \text{ odd} \end{cases}, \qquad \tilde{T}_i(x) = T_i(x) - T_0(x)
\end{equation}
where $T_i$ are the standard Chebyshev polynomials of the first kind.

This parameterization reduces the Boltzmann equation to an algebraic equation for the coefficients $\delta f^{a}_{ijk}$. Different from the fluid \emph{Ansatz} approach that integrates over phase space, this spectral method fixes these coefficients by demanding that the algebraic equation holds on a discrete grid of $(M-1)(N-1)^2$ points $\chi^{\alpha},\rho^{\beta}_z,\rho^{\gamma}_k$. These grid points are chosen as the roots of the Chebyshev polynomials:
\begin{equation}
\begin{aligned}
\chi^{(\alpha)} &= -\cos\left(\frac{\pi \alpha}{M}\right), \qquad &&\alpha = 1, \dots, M-1, \\
\rho^{(\beta)}_z &= -\cos\left(\frac{\pi \beta}{N}\right), \qquad &&\beta = 1, \dots, N-1, \\
\rho^{(\gamma)}_k &= -\cos\left(\frac{\pi \gamma}{N-1}\right), \qquad &&\gamma = 0, \dots, N-2.
\end{aligned}
\end{equation}
This choice of grid ensures the convergence of their method.

Evaluated on this grid, the Boltzmann equation simplifies into the matrix equation
\begin{equation}
\begin{aligned}
    \bigg( \partial_\xi \chi^{(\alpha)} \bigg[ &\mathcal{P}^{(\alpha, \beta, \gamma)}_w \partial_\chi - \frac{\gamma_w}{2} \partial_\chi(m^2)^{(\alpha)} (\partial_{p_z} \rho^{(\beta)}_z) \partial_{\rho_z} \bigg] \left[ \bar{T}_i(\chi^{(\alpha)}) \bar{T}_j(\rho^{(\beta)}_z) \tilde{T}_k(\rho^{(\gamma)}_k) \right] \delta_{ab} \\
    &+ \bar{T}_i(\chi^{(\alpha)}) C_{ab}^{\text{lin}} \left[ \bar{T}_j(\rho^{(\beta)}_z) \tilde{T}_k(\rho^{(\gamma)}_k) \right] \bigg) \delta f^b_{ijk} = \mathcal{S}_a \left( \chi^{(\alpha)}, \rho^{(\beta)}_z, \rho^{(\gamma)}_k \right),
    \label{eq:BE_wallgo}
\end{aligned}
\end{equation}
in the indices $\{\alpha,\beta,\gamma\}$ and ${i,j,k}$ and repeated indices are summed. 
Further details of the matrices and how the code works to find $\delta f^{b}_{ijk}$ in the equation~\eqref{eq:BE_wallgo} can be found in the \wallgo reference paper~\cite{Ekstedt:2024fyq}.

\section{Benchmarking velocities}
\label{sec:models}

In order to quantitatively evaluate the macroscopic hydrodynamical variables, one must specify the finite-temperature effective potential $\mathcal{F}(\phi,T)$.  We apply our methodology first to an extension of the Standard Model with a dimension-six operator (SMEFT), and then to a Standard Model-like theory (i.e. a purely quartic potential) with an artificially varying Higgs mass. By generating distinct thermodynamic profiles and transition strengths $\alpha$, these models allow us to explore a broader spectrum of hydrodynamical regimes. 

\subsection{The SMEFT model}
\label{subsec:smeft_benchmark}
We first adopt a Standard Model Effective Field Theory (SMEFT) including only a dimension-six operator. This scenario provides a robust mechanism for generating a very strong first-order electroweak phase transition while maintaining the physical Higgs mass at $m_h= 125$~GeV and the vacuum expectation value at $v = 246.22$~GeV.
The total finite-temperature effective potential is defined as

\begin{equation}
    \mathcal{F}(\phi,T) \equiv \mathcal{F}_0(\phi) + \sum_i (\pm g_i) \frac{T^4}{2\pi^2} \int_0^\infty dp \, p^2 \log\left( 1 \mp e^{-\sqrt{p^2 + (m_i(\phi)/T)^2}} \right).
    \label{eq:V_total_smeft}
\end{equation}
Here, the upper sign corresponds to bosons and the lower sign to fermions, $g_i$ is the number of degrees of freedom for each species, and $m_i(\phi)$ are the field-dependent masses. The zero-temperature potential $V_0(\phi)$ includes the tree-level self-interactions of the scalar field plus the one-loop Coleman-Weinberg corrections:
\begin{equation}
    \mathcal{F}_0(\phi)  = -\frac{\mu^2}{2}\phi^2 + \frac{\lambda}{4}\phi^4 + \frac{\phi^6}{8M^2} +\frac{\beta}{8} \left[ \phi^4 \left( \ln\left( \frac{\phi^2}{v^2} \right) - \frac{3}{2} \right) + 2\phi^2 v^2 \right].
\end{equation}
The $\phi^6$ operator encapsulates effects from unknown physics at an energy scale $M$, which we take as a free parameter of the model. The parameters $\mu^2$ and $\lambda$ are fixed so that this potential has its minimum in the vacuum expectation value $v$ at zero temperature, and that the second derivative of the potential at the minimum yields the known value for the squared Higgs mass. These conditions lead to
\begin{equation}
    \mu^2 = \frac{m_h^2}{2} - \frac{3v^4}{4M^2} \qquad \text{and} \qquad \lambda = \frac{m_h^2}{2v^2} - \frac{3v^2}{2M^2}.
\end{equation}
The top quark and gauge boson loops define the beta function 
\begin{equation}
    \beta = \frac{1}{8\pi^2 v^4} (6m_W^4 + 3m_Z^4 - 12m_t^4).
    \label{eq:beta}
\end{equation}

\subsection{The Light Higgs model}
As a complementary framework we adopt the light Higgs benchmark, where a first-order phase transition is generated by artificially lowering the Higgs mass to values $m_h \lesssim 80$~GeV within the Standard Model content. In this scenario, we evaluate the potential  using a high-temperature expansion~\cite{Moore_1995},
\begin{equation}
    \mathcal{F}(\phi, T) = D(T^2 - T_0^2)\phi^2 - e_T T |\phi|^3 + \frac{\lambda_T}{4}\phi^4 - c_T T^2 \phi^2 \ln\left(\frac{|\phi|}{T}\right) - \frac{\pi^2}{90}g_* T^4.
\end{equation}
The final term accounts for the radiation bath of the relativistic degrees of freedom in the symmetric phase, with $g_* = 28 + \frac{7}{8} (90) = 106.75$. To strictly mimic the Standard Model effective potential found in~\cite{Moore_1995, Ekstedt:2024fyq}, the parameters of the high-temperature expansion are defined as follows. The mass parameter $D$ is given by
\begin{equation}
    D = \frac{1}{8v^2} \left( 2m_W^2 + m_Z^2 + 2m_t^2 \right).
\end{equation}
The running thermal coupling $\lambda_T$ is defined as
\begin{equation}
    \lambda_T = \lambda_0 - \frac{3}{16\pi^2 v^4} \left[ 2m_W^4 \ln\left(\frac{m_W^2}{a_b T^2}\right) + m_Z^4 \ln\left(\frac{m_Z^2}{a_b T^2}\right) - 4m_t^4 \ln\left(\frac{m_t^2}{a_f T^2}\right) \right],
\end{equation}
where $\lambda_0 = m_h^2/(2v^2)$ is the tree-level coupling and the constants are $a_b \approx 49.78$ and $a_f \approx 3.11$.
The cubic term $e_T$ and the logarithmic coefficient $c_T$ are explicitly given by
\begin{equation}
    e_T = E_0 + \frac{3 + 3\sqrt{3}}{12\pi} |\lambda_T|^{3/2}, \qquad c_T = C_0 + \frac{1}{16\pi^2} \left( 4.8 g_2^2 \lambda_T - 6\lambda_T^2 \right),
\end{equation}
with $E_0 = \frac{1}{12\pi v^3} \left( 4m_W^3 + 2m_Z^3 \right)$, $C_0 \approx 1.42 g_2^4 / (16\pi^2)$, and $g_2 = 2m_W/v$.
Finally, the temperature parameter $T_0^2$ evaluates to
\begin{equation}
    T_0^2 = \frac{1}{4D} \left( m_h^2 - \beta v^2 \right)
\end{equation}
with $\beta$ given in eq.~\eqref{eq:beta} above.

\subsection{Results for $v_w$: \wallgo and the Fluid \emph{Ansatz}}
\label{sec:results}

\begin{figure}[t]
    \centering
    \begin{subfigure}[b]{0.45\textwidth}
        \centering
        \includegraphics[width=\textwidth]{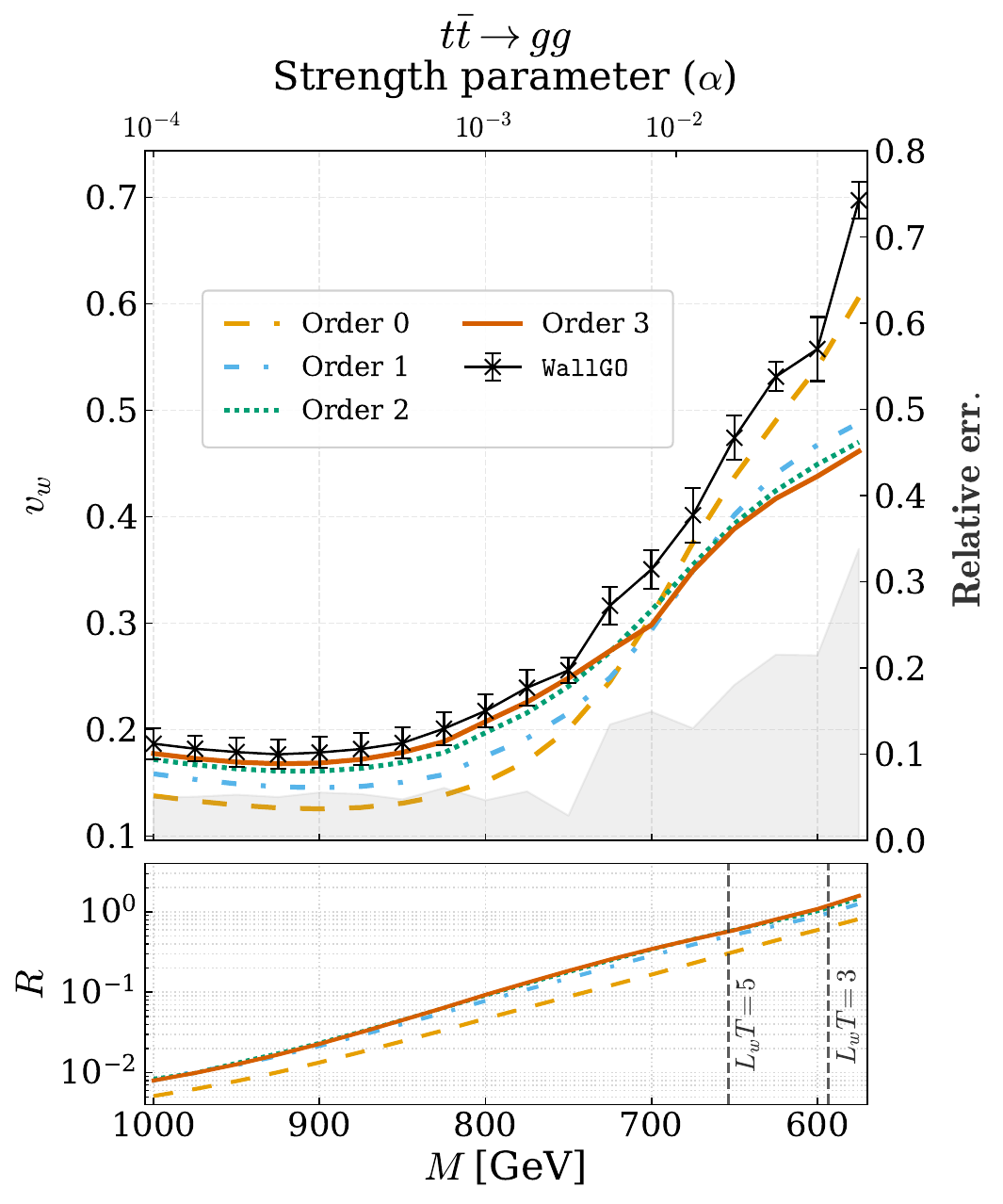}
        \caption{SMEFT Model}
        \label{fig:ttgg_smeft}
    \end{subfigure}
    \hfill
    \begin{subfigure}[b]{0.45\textwidth}
        \centering
        \includegraphics[width=\textwidth]{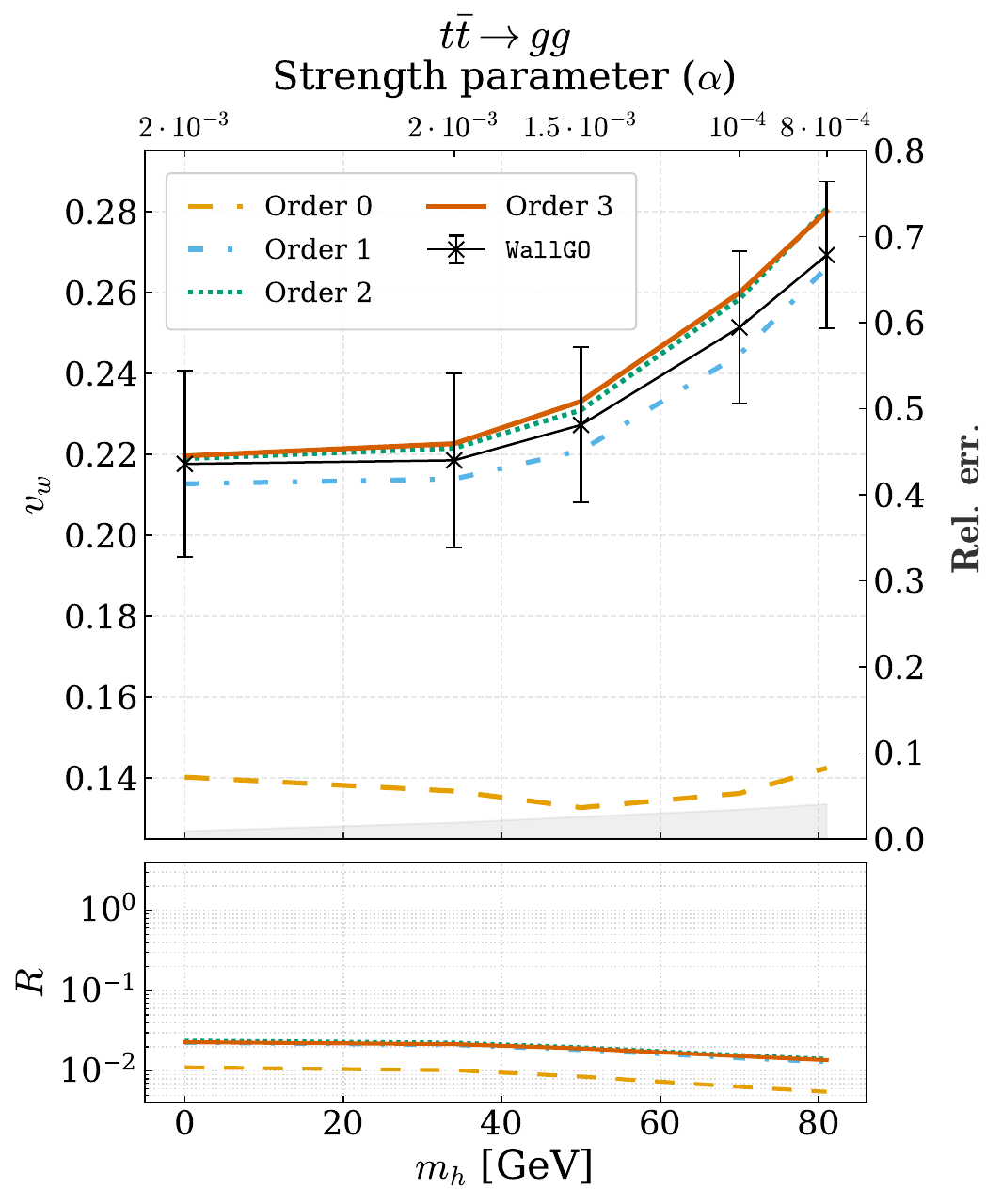}
        \caption{Light Higgs Model}
    \end{subfigure}
    \caption{Terminal wall velocity $v_w$ computed considering exclusively the $t\bar{t} \rightarrow gg$ processes. The left panel (a) shows the SMEFT benchmark as a function of the cutoff scale $M$, while the right panel (b) displays the Light Higgs scenario as a function of the Higgs mass $m_h$. The solution from the extended fluid \emph{Ansatz} at different truncation orders in momentum expansion is compared against the spectral method of \wallgo. The relative error of the third-order fluid \emph{Ansatz} with respect to the \wallgo result is plotted as the solid region, with numbers shown in the right axis of both figures. Additionally, the maximum residual $max|R|$ and the strength parameter of the transition as a function of the free parameter ($M$ or $m_h$) are also shown.}
    \label{fig:ttgg_comparison}
\end{figure}

\begin{figure}[h!]
    \centering
    \begin{subfigure}[b]{0.45\textwidth}
        \centering
        \includegraphics[width=\textwidth]{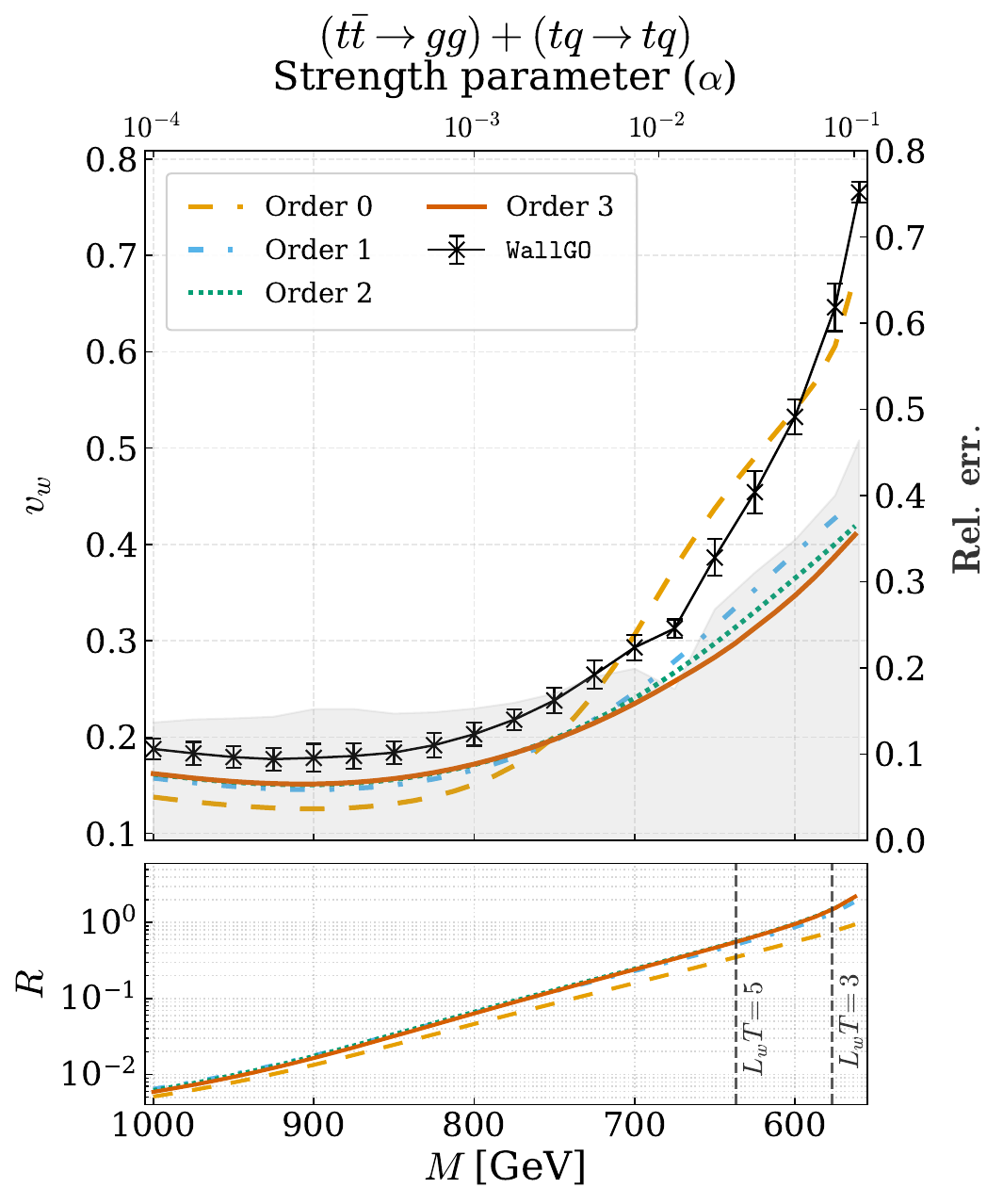}
        \caption{SMEFT Model}
        \label{fig:tgtq_combined_smeft}
    \end{subfigure}
    \hfill
    \begin{subfigure}[b]{0.45\textwidth}
        \centering
        \includegraphics[width=\textwidth]{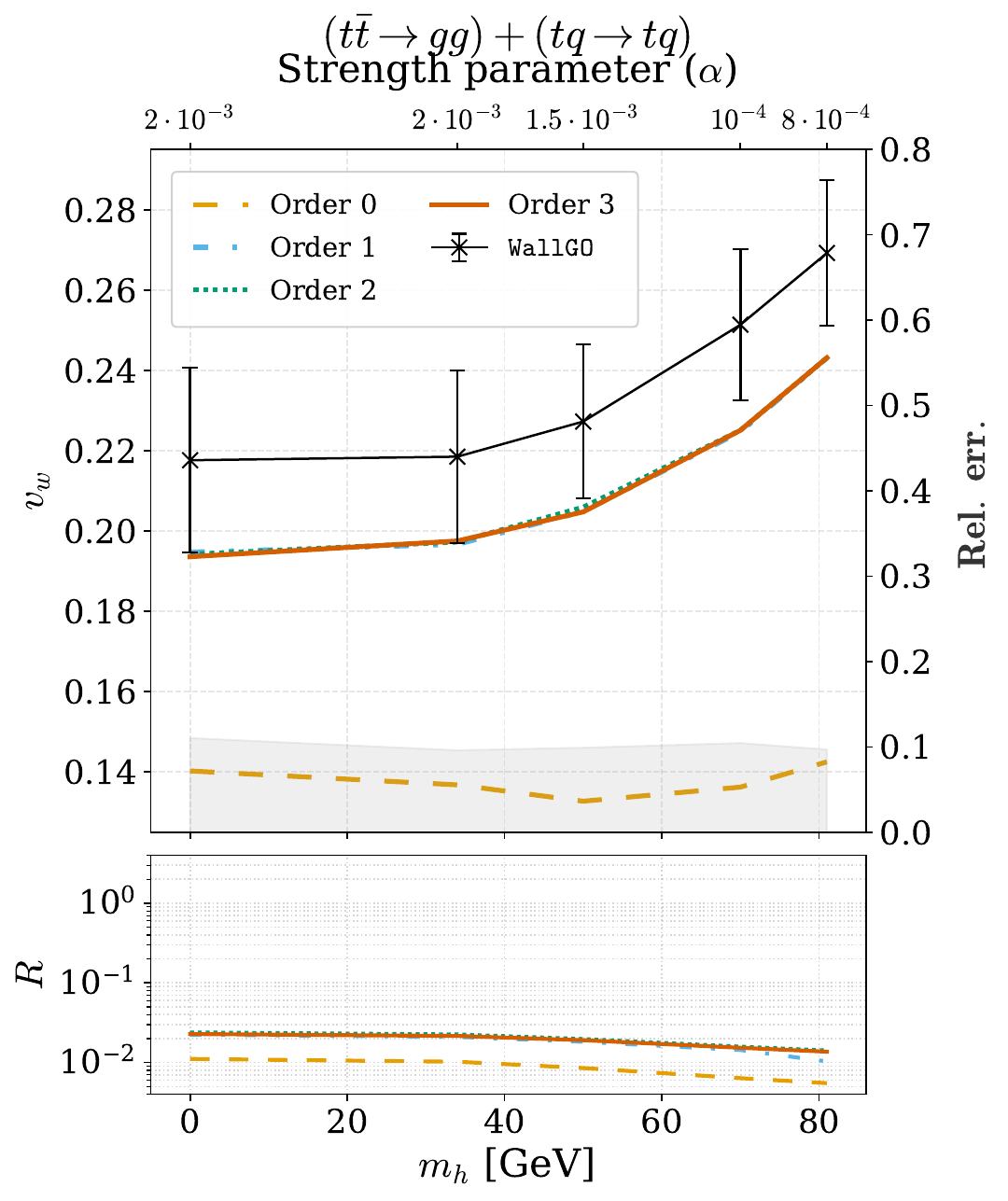}
        \caption{Light Higgs Model}
        \label{fig:tgtq_combined_light}
    \end{subfigure}
    \caption{Same as figure~\ref{fig:ttgg_comparison}, but computing the terminal wall velocity $v_w$ by considering the $tg \rightarrow tq$ and $t\bar{t}\rightarrow gg$ processes.}
    \label{fig:tgtq_combined_comparison}
\end{figure}

\begin{figure}[h!]
    \centering
    \begin{subfigure}[b]{0.45\textwidth}
        \centering
        \includegraphics[width=\textwidth]{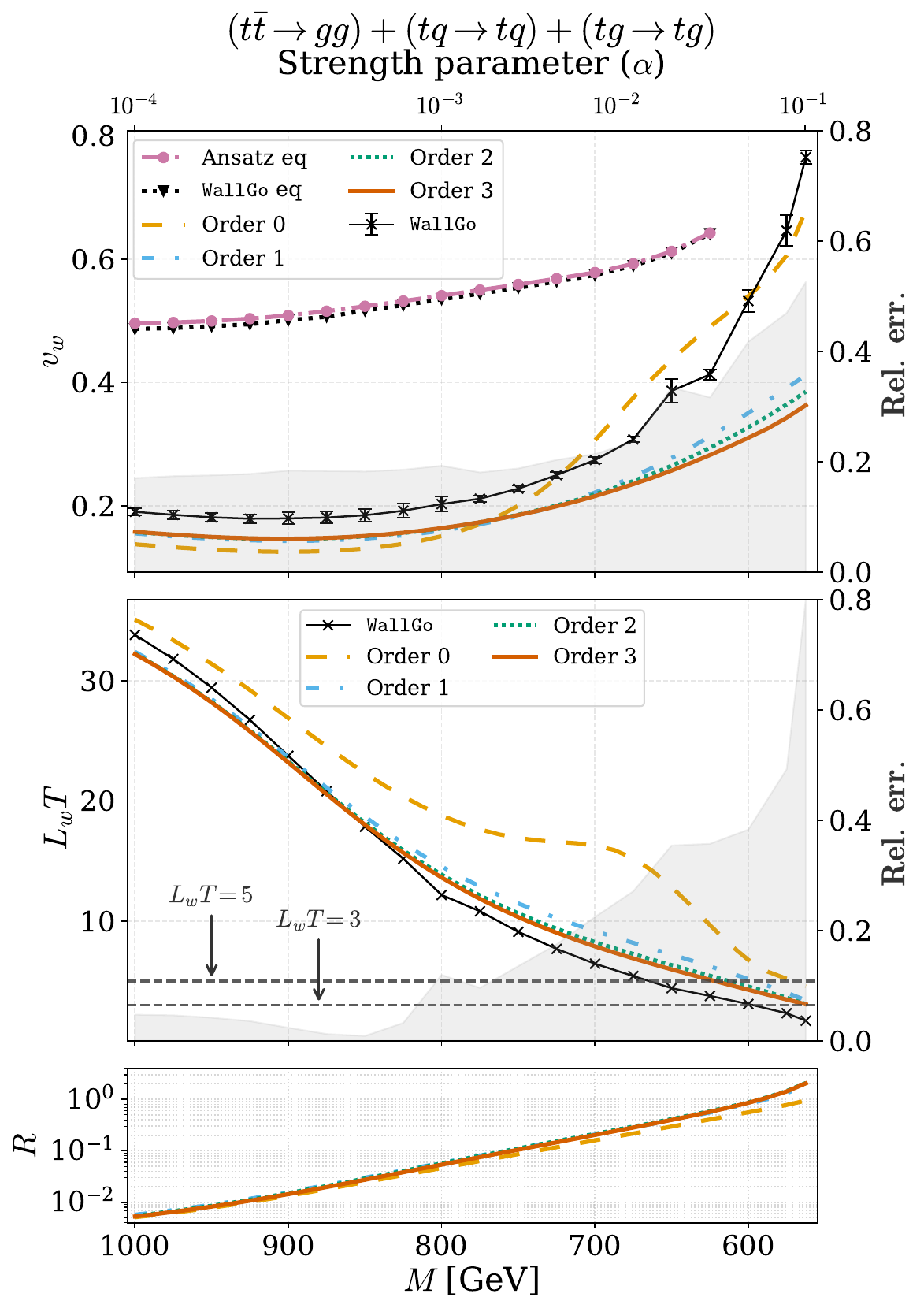}
        \caption{SMEFT Model}
        \label{fig:all_processes_smeft}
    \end{subfigure}
    \hfill
    \begin{subfigure}[b]{0.45\textwidth}
        \centering
        \includegraphics[width=\textwidth]{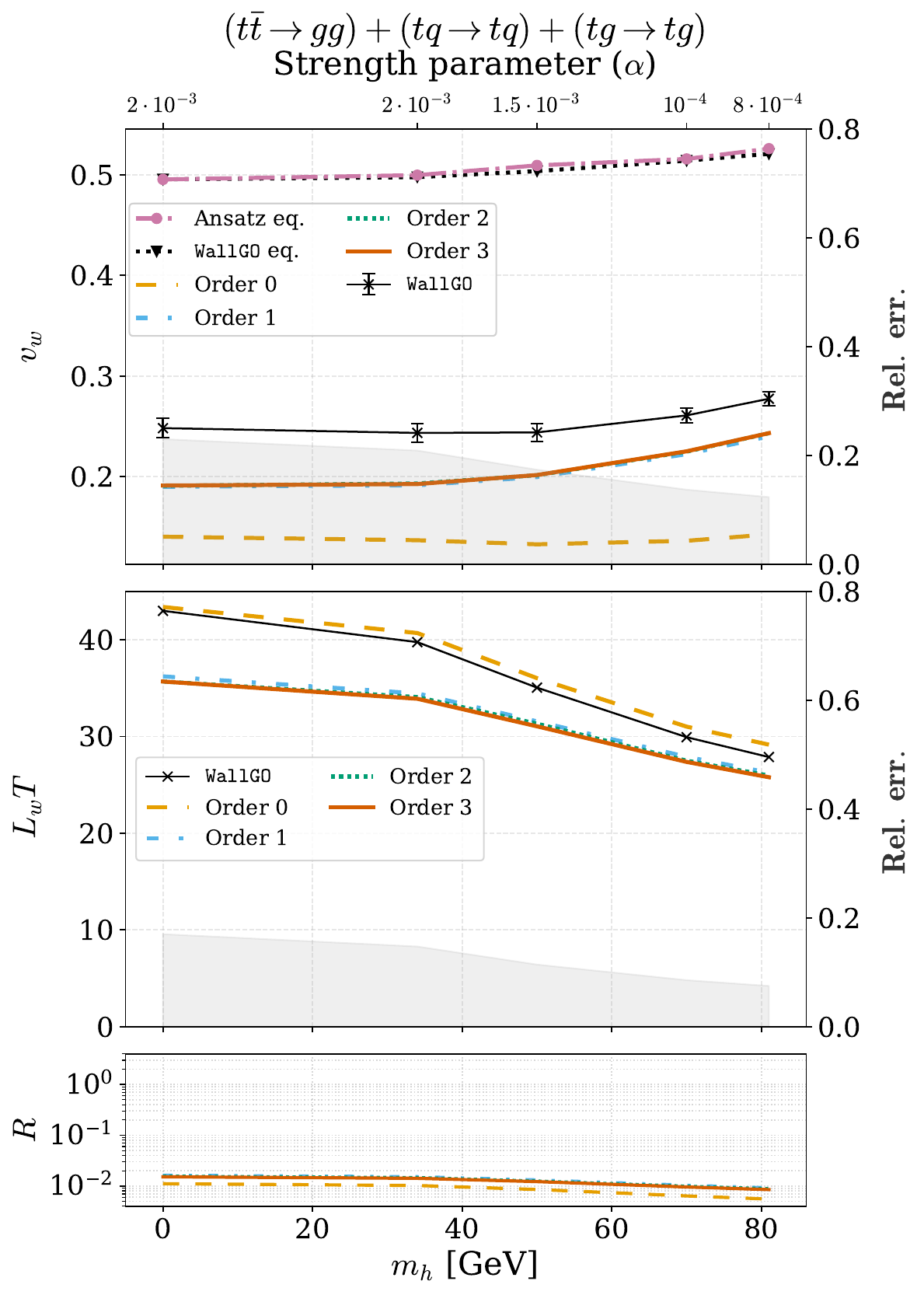}
        \caption{Light Higgs Model}
        \label{fig:all_processes_light}
    \end{subfigure}
    \caption{Total terminal velocity $v_w$ and the bubble wall thickness $L_w$ computed considering the combined effect of all leading-log top quark scattering processes ($t\bar{t}\rightarrow gg$, $tq \rightarrow tq$ and $tg\rightarrow$tg). The left panel (a) corresponds to the SMEFT solutions as a function of the cutoff scale $M$, while the right panel (b) shows the Light Higgs scenario as a function of the Higgs mass $m_h$. Solutions from the extended fluid \emph{Ansatz} evaluated at different truncation orders in momentum expansion are compared against the spectral method of \wallgo. The macroscopic wall velocities derived under the LTE are included for comparison for both methods. Following the layout of figure~\ref{fig:ttgg_comparison}, the relative error of the third fluid \emph{Ansatz} with the \wallgo is plotted as the solid region on the right axis, and the maximum residual $\max |R|$ along with the transition strength parameter are shown for each respective model.}
    \label{fig:all_processes_comparison}
\end{figure}

We implemented both models above in the \wallgo package, which is already fully automated to calculate the bubble wall velocity given the free-energy, following the prescriptions detailed in ref.~\cite{Ekstedt:2024fyq}. Note that the \wallgo code also automatically computes the collision matrix under the assumption of an expansion in Chebyshev polynomials --- the user only needs to specify which processes are to be taken into account. For the fluid \emph{Ansatz}, we use a custom-made \texttt{Python} code for solving the Boltzmann equation via Green's method as described in section~\ref{sec:fluidAnsatz} and then solving the two moments of the Higgs equation-of-motion for $v_w$ and $L_w$ as shown in section~\ref{sec:KGeq}. The collision matrix for the fluid \emph{Ansatz} is computed according to the leading-logarithm prescription outlined in Appendix~\ref{sec:app_collision}. Below we show results for the extended fluid \emph{Ansatz} considering up to order 3 in momentum expansion, which amounts to up to 10 non-equilibrium fluctuations. For the \wallgo results presented in this section we set the momentum and spatial grid sizes to $ N =11$ and $M_{\text{grid}} =30$, respectively. The integration options and integral solver parameters (\texttt{temperatureVariationScale, meanFreePathScale, fieldValueVariationScale}) were kept at the standard default configurations provided by the code for the Light Higgs model. This parameter choice demonstrated robust convergence, while modifying or increasing the grid resolution and solver scales yielded no significant deviations, with variations in the terminal velocity bounded to a maximum of 2\% relative to the baseline configuration. In the bottom panels of figures~\ref{fig:ttgg_comparison},~\ref{fig:tgtq_combined_comparison} and~\ref{fig:all_processes_comparison} we show the perturbativity parameter $R$ defined in eq.~\eqref{eq:R}.

To benchmark the non-equilibrium modeling techniques, we first isolate the macroscopic impact of the annihilation channel. Figure~\ref{fig:ttgg_comparison} presents the terminal wall velocity when including only the $t\bar{t} \to gg$ annihilation process. For the fluid \emph{Ansatz}, results are shown for various truncations in momentum expansion, up to order 3 (amounting to 10 fluctuations). We first note that the results for the fluid \emph{Ansatz} converge to a unique specific value when more fluctuations are included, showing that the expansion of the non-equilibrium parameter $\delta$ in powers of momenta is well behaved.  Moreover, for both the SMEFT and the Light Higgs scenarios we find a robust agreement between the truncated fluid \emph{Ansatz} at order 3 and the \wallgo spectral method across a wide range of transition strengths. Note that here and in the other figures the disagreement between results from both approaches start to be noticeable precisely when the perturbativity parameter approaches its limiting value $R\approx 1$, which already points to a breakdown of the linearization procedure performed in eq.~\eqref{eq:fluid_delta}. We will discuss in section~\ref{sec:second_order} the impact of including the quadratic terms as well. Although seemingly pointing to a limitation of the linearized fluid \emph{Ansatz}, it is also worth noticing that in this range the wall thickness is also approaching its limiting value $L_w T\to 1$, as also indicated in the lower panels, thus pointing to a breakdown of the whole WKB approximation that led to the Boltzmann equation. So it is overall unclear whether any strong conclusions can be drawn from the results in this region.

Figure~\ref{fig:tgtq_combined_comparison} illustrates the wall velocity prediction when both $t\bar{t}\rightarrow gg$ (top annihilation) and $tq\rightarrow tq$ (top scattering off quarks) are simultaneously taken into account. Here one can appreciate an advantage of the fluid \emph{Ansatz}, namely the physical intuition attributable to each of the fluctuations. As discussed in section~\ref{sec:fluidAnsatz}, the ``order 0'' curve considers only a momentum-independent fluctuation, which can be interpreted as the chemical potential. This fluctuation is unaffected by scatterings (since these processes do not alter particle number) and therefore the curve is the same as for figure~\ref{fig:ttgg_comparison}. At ``order 1'' one includes temperature and velocity fluctuations as well, but now the presence of scattering processes will help equilibrate these fluctuations faster than if only annihilations were present. Indeed, one observes that, at least for sufficiently weak phase transitions ($\alpha \lesssim 10^{-2}$), the higher order curves are closer to the ``order 0'' curve compared to the case of annihilations only (compare with figure~\ref{fig:ttgg_comparison}). For stronger transitions the overall trend that scatterings damp the thermal and velocity fluctuations remains valid, but the resulting effect on the wall velocity is less trivial: the thermal fluctuations diminish, but since they actually provided a \emph{negative} contribution to friction this means the overall friction increases and the wall velocity is thus reduced. Still, as a sanity check, we have verified that if we multiply the scattering collision terms by an arbitrarily large factor the curve would indeed converge to the ``order 0'' result in the entire range of parameter space, as intuitively expected.

Finally, figure~\ref{fig:all_processes_comparison} shows the dependence of the wall velocity and the wall thickness with the free parameter of the underlying model, now considering the total combined effect of taking into account top annihilations and scatterings off quarks and off gluons. In this figure we also show the velocities (calculated both by \wallgo and by our custom-made code) considering only the equilibrium contribution to the friction (coming from the variation of the background fluid temperature~\cite{Dorsch:2024jjl}). The fact that both curves agree to an excellent precision provides us with a good sanity check that our implementation of the models is indeed equivalent in both cases. We successfully checked that, if the collision terms were all multiplied by an arbitrarily large factor (meaning extremely efficient equilibration processes), all the curves would agree with this case of ``equilibrium friction only'', as they should. 

Our comparative analysis reveals a strong correspondence for isolated processes, with precision decreasing upon the simultaneous inclusion of multiple channels, or particularly for strong phase transitions (lower $M$ in SMEFT). In these strong regimes, the terminal velocity predicted by the fluid \emph{Ansatz} starts to deviate from the \wallgo result. Concurrently, the maximum residual $\max|R|$ experiences a clear increase, signaling the generation of large macroscopic fluctuation in the plasma. These large fluctuations strongly indicate a breakdown of the truncated linear approximation $\mathcal{O}(\delta)$ within the fluid \emph{Ansatz} framework, as the departures from equilibrium become too large to be accurately captured by the first-order Taylor expansion.

Furthermore, it is important to emphasize a key point concerning the dimensionless wall thickness $L_w  T$. As the transition strength $\alpha$ increases, $L_w  T$ decreases monotonically. In the regions of our parameter space where $L_w T$ approaches unity, the WKB approximation underlying the derivation of the semi-classical Boltzmann equation~\eqref{eq. BEq} approaches the limits of its validity for both numerical approaches. At the same time, the increasingly large values of $\max |R|$ clearly indicate the need to investigate higher-order non-equilibrium effects, which we address in section~\ref{sec:second_order}. Throughout this analysis we work within the SMEFT framework, assuming that the top quark is the only particle species driven out of equilibrium and considering the $t\bar{t}\rightarrow gg$ scattering process exclusively.

 \subsection{Convergence behaviour with the addition of parameters}
 
In the above analysis we have fixed the number of parameters used in the \wallgo evaluation, and compared this prediction with those from the fluid \emph{Ansatz} for various transition strengths and various truncation orders in momentum expansion of the fluid approach. Now we would also like to check how well each \emph{Ansatz} performs for a fixed transition strength, but varying the number of parameters used in the expansion of the non-equilibrium particle distribution function in each approach. I.e. we quantify how many parameters each \emph{Ansatz} actually needs in order to reach a reasonably reliable prediction for $v_w$.

For \wallgo the precision of the solutions depends mainly on the resolution of the Chebyshev polynomials basis, which is governed by the number of spatial points $M_{\text{grid}}$ and momentum grid points $N$, generating $D = (M_{\text{grid}}-1)(N-1)^2$ basis parameters. The accuracy of the fluid \emph{Ansatz} relies intrinsically on the truncation order $\kappa$ of the momentum expansion for the non-equilibrium distribution function. Increasing $\kappa$ expands the system of differential equations to include $D = (\kappa+1)(\kappa+2)/2$ fluctuations.

\begin{figure}[t]
    \centering
    \includegraphics[width=0.7\textwidth]{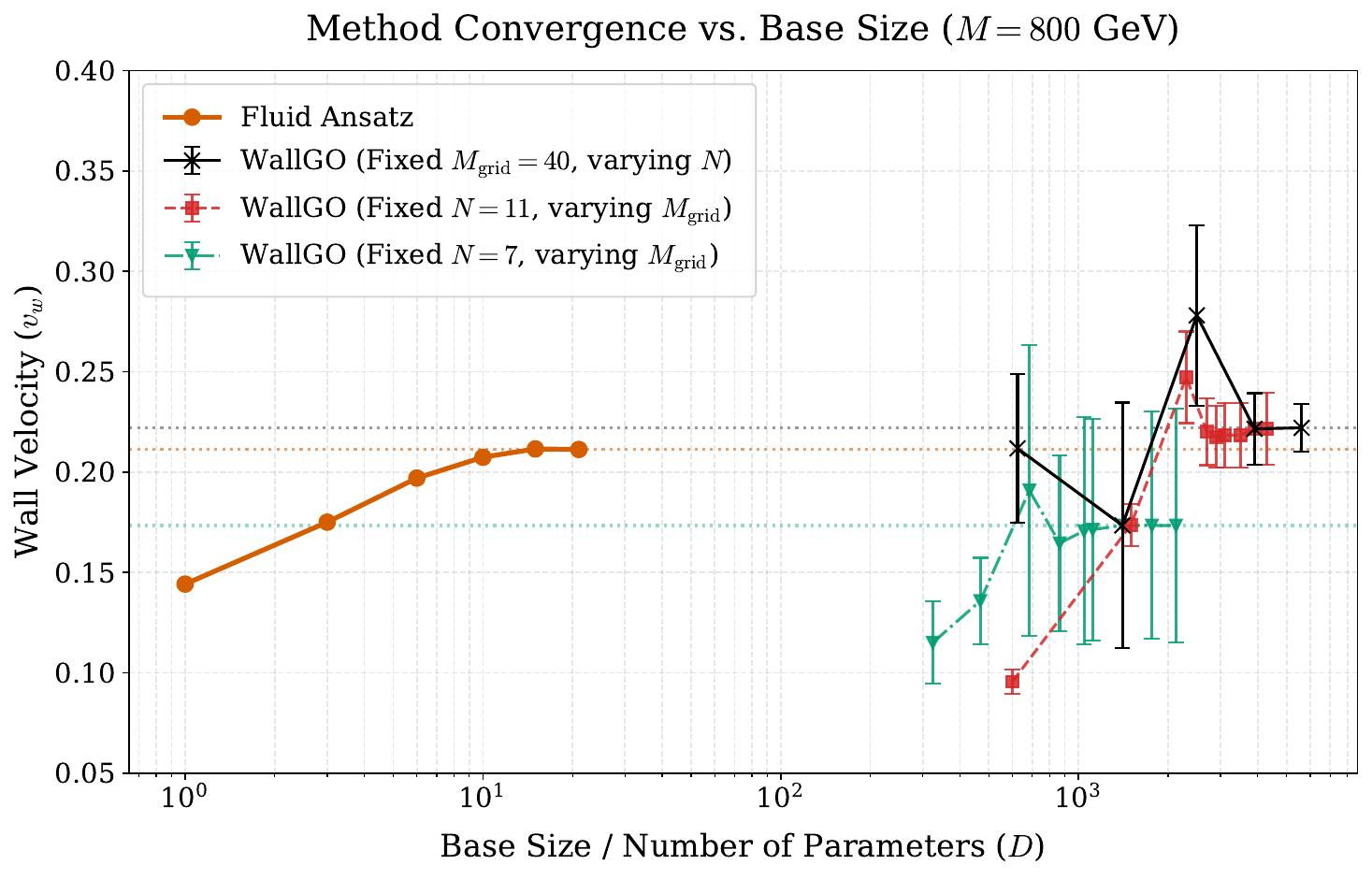}
    \hfill
    \caption{Convergence of the predicted wall velocity in the extended fluid \emph{Ansatz} and in the \wallgo approach as a function of the number of parameters involved in each expansion.}
    \label{fig:vw_convergence}
\end{figure}

We performed three different scans for the \wallgo method: varying the spatial grid $M_\text{grid}$ while keeping the momentum grid fixed at a lower resolution ($N=7$) and a higher resolution ($N =11$), as well as varying the momentum resolution $N$ while fixing $M_{\text{grid}}=40$. The results in figure~\ref{fig:vw_convergence} demonstrate that if the two variables $M_{\text{grid}}$ and $N$ are not sufficiently large, the number of basis coefficients fails to adequately capture the spatial profile ($\chi$) and the momentum profiles ($p_\parallel$ and $p_\perp$). This can be seen in the $N= 7$ scan, for which, despite increasing $M_{\text{grid}}$, the solution stabilizes at an inaccurate asymptotic velocity with large uncertainty due to lack of momentum resolution. Conversely, when the momentum grid is sufficiently resolved, the non-equilibrium dynamics stabilize. We also take the opportunity to observe that, in some cases, the \wallgo error bars are not reliable predictors of the correct terminal velocity, as they do not encompass the asymptotic solution finally achieved as one increases the number of fluctuations. 

On the other hand, for the fluid \emph{Ansatz} the solution shows good convergence with $D(\kappa =3) = 10$ parameters. Ultimately, when adequately resolved, both methods converge to asymptotic velocities that agree within a margin of $5\%$.

\section{$\mathcal{O}\left(\delta ^2\right)$ contributions}
\label{sec:second_order}

In figures~\ref{fig:ttgg_comparison}--\ref{fig:all_processes_comparison} we showed that the wall velocities computed with the fluid \emph{Ansatz} are in good agreement with \wallgo predictions for weaker transitions ($\alpha\lesssim 10^{-2}$). They start to diverge precisely when the fluctuations start to become large, with the perturbativity parameter $R\gtrsim 0.1$. This motivates us to investigate the impact of including non-linear effects associated to $\mathcal{O}(\delta^2)$ effects in the Taylor expansion of the distribution function.

\subsection{Second-order Kinetic Terms}

The non-equilibrium distribution function in the fluid \emph{Ansatz} is postulated to have the form $f = (e^{\beta(p^\mu u_\mu - \delta)} \pm 1)^{-1}$. Defining $f_{eq}' \equiv \partial f_{\text{eq}} / \partial(p^\mu u_\mu)$, a Taylor expansion yields
\begin{equation}
f \simeq f_{\text{eq}} +\underbrace{ \left(-f_{eq}' \delta + \frac{1}{2} f_{eq}'' \delta^2+\ldots\right)}_{\delta f}.
\label{eq:taylor}
\end{equation}
Let us investigate how the inclusion of the second-order terms $\mathcal{O}(\delta^2)$ affects the Boltzmann equation.

Applying this expansion to the Liouville operator, the Boltzmann equation naturally separates into a background source term and a kinetic term, and with the new expansion \ref{eq:taylor}, the Boltzmann equation can be written as
\begin{equation}
\begin{split}
    \mathcal{S}(p,\xi)+
    p^{\nu}\left[(-f_{eq}')\partial_\nu \delta + \delta\partial_\nu(-f_{eq}') +\delta (f_{eq}'')\partial_\nu \delta  \right] \\
    &\hspace{-100pt}+\partial_\nu \left( \frac{m
    _i^2}{2} \right) \left[ - f_{eq}' \partial_{p_\nu} (\delta) - \delta \beta (-f_{eq}'')u^\nu \right]
     = -\mathcal{C}[\delta f]    
     \label{eq:new_BE}
\end{split}
\end{equation}
To construct a tractable system of differential equations, the momentum expansion of the non-equilibrium perturbation in eq.~\eqref{eq:delta} is truncated at first order. Under this expansion, the departure from equilibrium is entirely parametrized by the fluid variables: the chemical potential $\delta \mu$, the temperature fluctuation $\delta T$, and the fluid velocity perturbation $\delta v$, explicitly
\begin{equation}
    \delta(\xi,p) = \delta \mu + p^\mu \left( \delta u_\mu - u_\mu \frac{\delta T}{T} \right).
    \label{eq:truncation}
\end{equation}
By decomposing the kinetic operator of eq.~\eqref{eq:new_BE} into a drift operator  $ \mathcal{D}\equiv p^\nu\partial_\nu $ and a force operator $F\equiv \partial_\nu\frac{m_i^2}{2}\partial_{p_{\nu}}$we can explicitly separate the total kinetic contribution as $\mathcal{K}[(\delta f)]\equiv \mathcal{D}[(\delta f)]+F[(\delta f)] $ .Truncating the momentum expansion at order $\kappa =1$, and using the defined constants of eq.~\eqref{eq:constants_c}, this yields a closed system for $D = 3$ macroscopic moments for the Boltzmann equation. Therefore, integrating over the phase-space and taking these 3 moments from eq.~\eqref{eq:new_BE} leads to
\begin{equation}
\begin{aligned}
    \mathcal{D}^1 &= \frac{\gamma_w}{2\pi^2} \Bigg[ \mu^\prime \left( v_w c_2 - v_w \kappa_2 \mu - v_w \kappa_3 \delta T - \frac{\kappa_3}{3} \delta v \right) \\
    &\quad + \delta T^\prime \left( v_w c_3 - v_w \kappa_3 \mu - v_w \kappa_4 \delta T - \frac{\kappa_4}{3} \delta v \right) \\
    &\quad + \delta v^\prime \left( \frac{c_3}{3} - \frac{\kappa_3}{3} \mu - \frac{\kappa_4}{3} \delta T - v_w \frac{\kappa_4}{3} \delta v \right) \\
    &\quad - (\kappa_3 \mu + \kappa_4 \delta T) \left( v_w \frac{\partial_\xi T}{T} - \frac{\gamma_w^2 \partial_\xi V_{pl}}{3} \right) - \frac{\kappa_4}{3} \delta v \left( \frac{\partial_\xi T}{T} - v_w \gamma_w^2 \partial_\xi V_{pl} \right) \Bigg],
\end{aligned}
\end{equation}
\begin{equation}
\begin{aligned}
    \mathcal{D}^2 &= \frac{\gamma_w}{2\pi^2} \Bigg[ \mu^\prime \left( v_w c_3 - v_w \kappa_3 \mu - v_w \kappa_4 \delta T - \frac{\kappa_4}{3} \delta v \right) \\
    &\quad + \delta T^\prime \left( v_w c_4 - v_w \kappa_4 \mu - v_w \kappa_5 \delta T - \frac{\kappa_5}{3} \delta v \right) \\
    &\quad + \delta v^\prime \left( \frac{c_4}{3} - \frac{\kappa_4}{3} \mu - \frac{\kappa_5}{3} \delta T - v_w \frac{\kappa_5}{3} \delta v \right) \\
    &\quad - (\kappa_4 \mu + \kappa_5 \delta T) \left( v_w \frac{\partial_\xi T}{T} - \frac{\gamma_w^2 \partial_\xi V_{pl}}{3} \right) - \frac{\kappa_5}{3} \delta v \left( \frac{\partial_\xi T}{T} - v_w \gamma_w^2 \partial_\xi V_{pl} \right) \Bigg],
\end{aligned}
\end{equation}
\begin{equation}
\begin{aligned}
    \mathcal{D}^3 &= \frac{\gamma_w}{2\pi^2} \Bigg[ \mu^\prime \left( \frac{c_3}{3} - \frac{\kappa_3}{3} \mu - \frac{\kappa_4}{3} \delta T - v_w \frac{\kappa_4}{3} \delta v \right) \\
    &\quad + \delta T^\prime \left( \frac{c_4}{3} - \frac{\kappa_4}{3} \mu - \frac{\kappa_5}{3} \delta T - v_w \frac{\kappa_5}{3} \delta v \right) \\
    &\quad + \delta v^\prime \left( v_w \frac{c_4}{3} - v_w \frac{\kappa_4}{3} \mu - v_w \frac{\kappa_5}{3} \delta T - \frac{\kappa_5}{5} \delta v \right) \\
    &\quad - \frac{1}{3} (\kappa_4 \mu + \kappa_5 \delta T) \left( \frac{\partial_\xi T}{T} - v_w \gamma_w^2 \partial_\xi V_{pl} \right) - \kappa_5 \delta v \left( \frac{v_w}{3} \frac{\partial_\xi T}{T} - \frac{\gamma_w^2 \partial_\xi V_{pl}}{5} \right) \Bigg].
\end{aligned}
\end{equation}
Similarly, the mass-gradient force components $\mathcal{F}^i$ can be grouped like
\begin{equation}
\begin{aligned}
    F^1 &= \frac{\gamma_w}{(2\pi^2)}  \Big[ v_w \kappa_1 \mu + v_w (c_1 + \kappa_2) \delta T + c_1 \delta v \Big] \partial_\xi \frac{m_i^2}{2T^2}, \\
    F^2 &= \frac{\gamma_w}{(2\pi^2)}\Big[ v_w \kappa_2 \mu + v_w (c_2 + \kappa_3) \delta T + c_2 \delta v \Big]\partial_\xi \frac{m_i^2}{2T^2}, \\
    F^3 &=  \frac{\gamma_w}{(2\pi^2)} \Big[ v_w \frac{\kappa_3}{3} \delta v \Big]\partial_\xi \frac{m_i^2}{2T^2}.
\end{aligned}
\end{equation}

\subsection{Second-Order Collision Terms}
The collision operator must correspondingly be expanded to $\mathcal{O}(\delta^2)$. Evaluating the population factor reveals an exact cancellation of the isolated quadratic terms, $\Lambda_i \equiv \frac{1}{2}(\pm f_{i}^{eq})(1 \pm f_{i}^{eq})\delta_i^2$, between the forward and inverse processes. Therefore, it can be shown that the net second-order collision rate factors neatly into the standard linear term scaled by a nonlinear factor
\begin{equation}
 \mathcal{P}^{(2)} = f_p^{eq}f_k^{eq}(1+s_{p'}f_{p'}^{eq})(1+s_{k'}f_{k'}^{eq}) (\delta_p + \delta_k - \delta_{p^\prime} - \delta_{k^\prime})  \left[ 1 + \sum_{i \in \{p,k,p^\prime,k^\prime\}} \left(\frac{1}{2} + s_i f_{i}^{eq}\right) \delta_i \right],
\end{equation}
where $s_i = +1$ for bosons and $-1$ for fermions.

With this derivation, one can recycle the definition of the constant $\gamma^{pk \to p^\prime k^\prime}_{mnrs}$ related to linear collision terms to construct the second-order dimensionless constant
\begin{equation}
\begin{split}
 \omega^{pk \to p^\prime k^\prime}_{mnrstl} &= \frac{1}{T^{m+n+s+r+4}}\frac{1}{16(2\pi)^8} \int \frac{d^3p\, d^3k}{E_p\, E_k} f_{p}^{\text{eq}} f_{k}^{\text{eq}} \\
 &\quad \times \int \frac{d^3p^\prime\, d^3k^\prime}{E_{p^\prime}\, E_{k^\prime}} \bigl(1 + s_{p^\prime} f_{p^\prime}^{\text{eq}}\bigr)\bigl(1 + s_{k^\prime} f_{k^\prime}^{\text{eq}}\bigr) |\mathcal{M}|^2 \delta^{(4)}(p + k - p^\prime - k^\prime) \\
&\hspace{40pt}\quad \times (p^\mu u_\mu)^{m} (p^\nu \bar{u}_\nu)^{r} \Big( h_p(p^{\mu}u_{\mu})^n(p^{\rho}\bar{u}_{\rho})^s+h_{k}(k^{\mu}u_{\mu})^n(k^{\rho}\bar{u}_{\rho})^s\\
&\hspace{120pt}-h_{p'}(p'^{\mu}u_{\mu})^n(p'^{\rho}\bar{u}_{\rho})^s-h_{k'}(k'^{\mu}u_{\mu})^n(k'^{\rho}\bar{u}_{\rho})^s\Big) \\
&\hspace{90pt}\times \left[ \hspace{15pt} \hspace{-10pt}\sum_{\sigma \in \{p,k,p^\prime,k^\prime\}} h_\sigma\left(\frac{1}{2} + s_\sigma f_{\sigma}^{eq}\right) \left(\frac{\sigma^\mu u_\mu}{T}\right)^t\left(\frac{\sigma^\mu \bar u_\mu}{T}\right)^l\right].
\end{split}
\end{equation}
The projection powers $m$ and $r$ tell us the specific moment being evaluated. These correspond to the momentum powers $a_i,b_i$ defined at the beginning of Appendix~\ref{app:sources_en_mom}. Therefore, for the $i$-th moment of the Boltzmann equation, we set $m=a_i$ and $r=b_i$. It's also important to stress that the internal index pairs $(n,s)$ and $(t,l)$ map to the momentum weights of the interacting non-equilibrium variables. Let us define the non-equilibrium vector $x=( x_1,x_2,x_3)\equiv (\mu, \delta T, \delta v)$. We assign each macroscopic variable $x_A$ to its corresponding basis powers $(n_A,s_A)$, where $n_A$ is associated with the projection $p^{\mu}u_\mu$ and $s_A$ associated with the projection $p^{\mu}\bar u_\mu$. Therefore, one can write the new collision term associated with the $i$-th Boltzmann equation's moment as
\begin{equation}
    \hspace{-5pt}\Gamma^{i}_{\text{new}} = \Gamma^{ij}_{\text{old}}q_j +\Gamma^{i}_{{\mathcal{O}(\delta^2)}} =\sum_{\text{processes}}\Big [\sum_{A=1}^{3} x_A \gamma^{pk \to p^\prime k^\prime}_{a_i,n_A,b_i,s_A}+  \sum_{A=1}^{3}\sum_{B=1}^3 x_A x_B \omega^{pk \to p^\prime k^\prime}_{a_i,n_A,b_i,s_A,n_B,s_B}\Big]
\end{equation}

\subsection{Solving the non-linear System}
Substituting the macroscopic variables $\mu, \delta T$, and $\delta v$, the complete system of equations is given by
\begin{equation}
  \mathcal{D}^i(q,q',\xi) +F^{i}(q,\xi) + (L_w T)\Gamma_{\text{new}}^i(q,\xi) = S^i(\xi), \qquad i = {1,2,3},
\end{equation}
where the formula for the source terms $\mathcal{S}^{i}(z)$ remains identical to the linear case defined in eq.~\eqref{eq:Source}.

Unlike the linear approximation in the Taylor expansion, the inclusion of the $\mathcal{O}(\delta^2)$ contributions leads to a highly non-linear system of equations. Solving this Boundary Value Problem (BVP) across the extended wall profile is a significant numerical challenge due to the inherent stiffness introduced by the quadratic perturbations. To reliably extract the solutions, the non-linear BVP is solved iteratively via collocation methods. To ensure numerical stability and proper convergence, the solution via Green's method in eq.~\eqref{eq:q_fluc} is provided as the initial cross-check for the non-linear solver.

\begin{figure}[H]
    \centering
    \begin{subfigure}{0.48\textwidth}
    \centering
    \includegraphics[height=10cm,keepaspectratio]{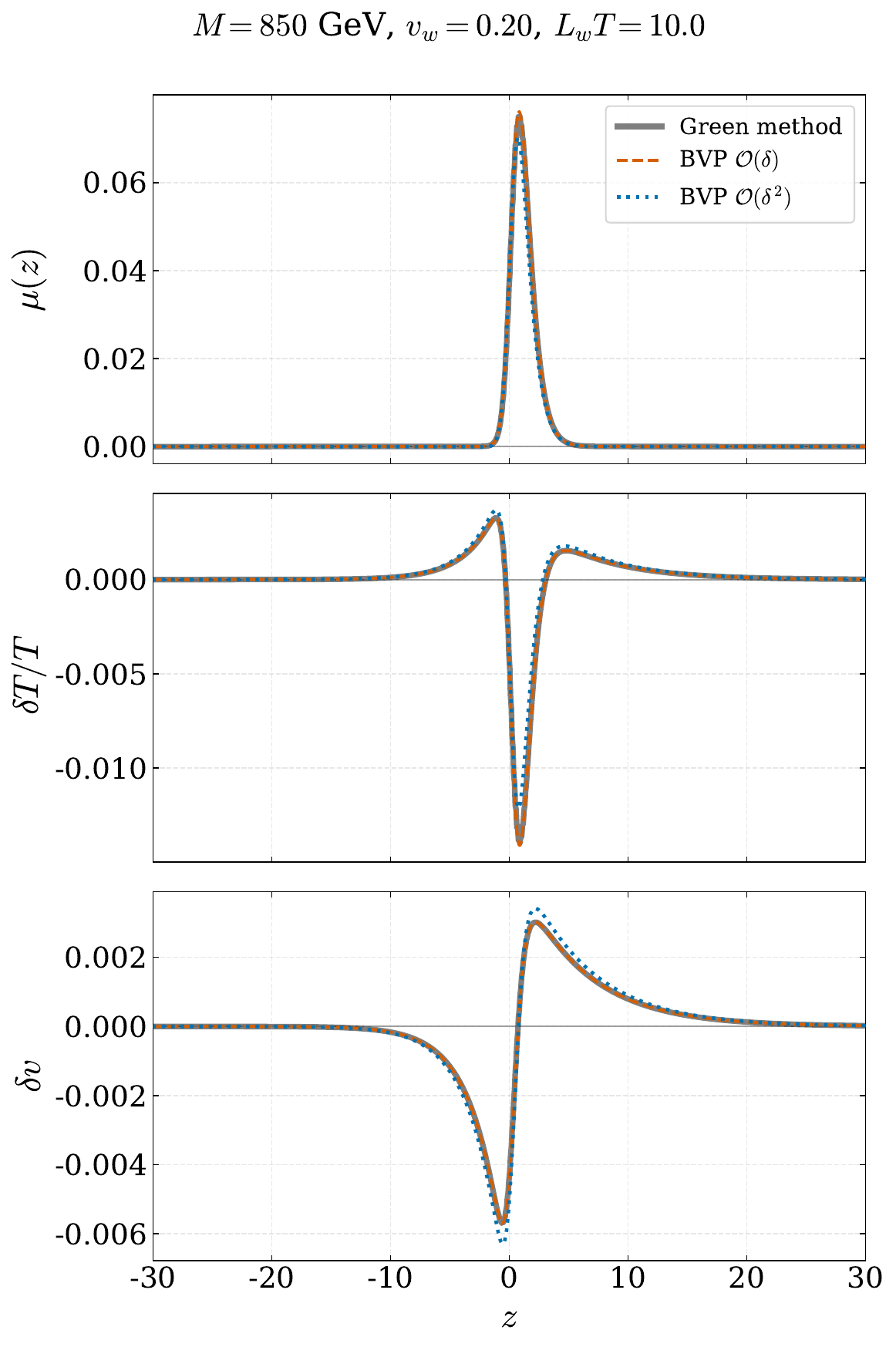}
    \end{subfigure}
    \hfill
    \begin{subfigure}{0.48\textwidth}
    \centering
    \includegraphics[height=10cm,keepaspectratio]{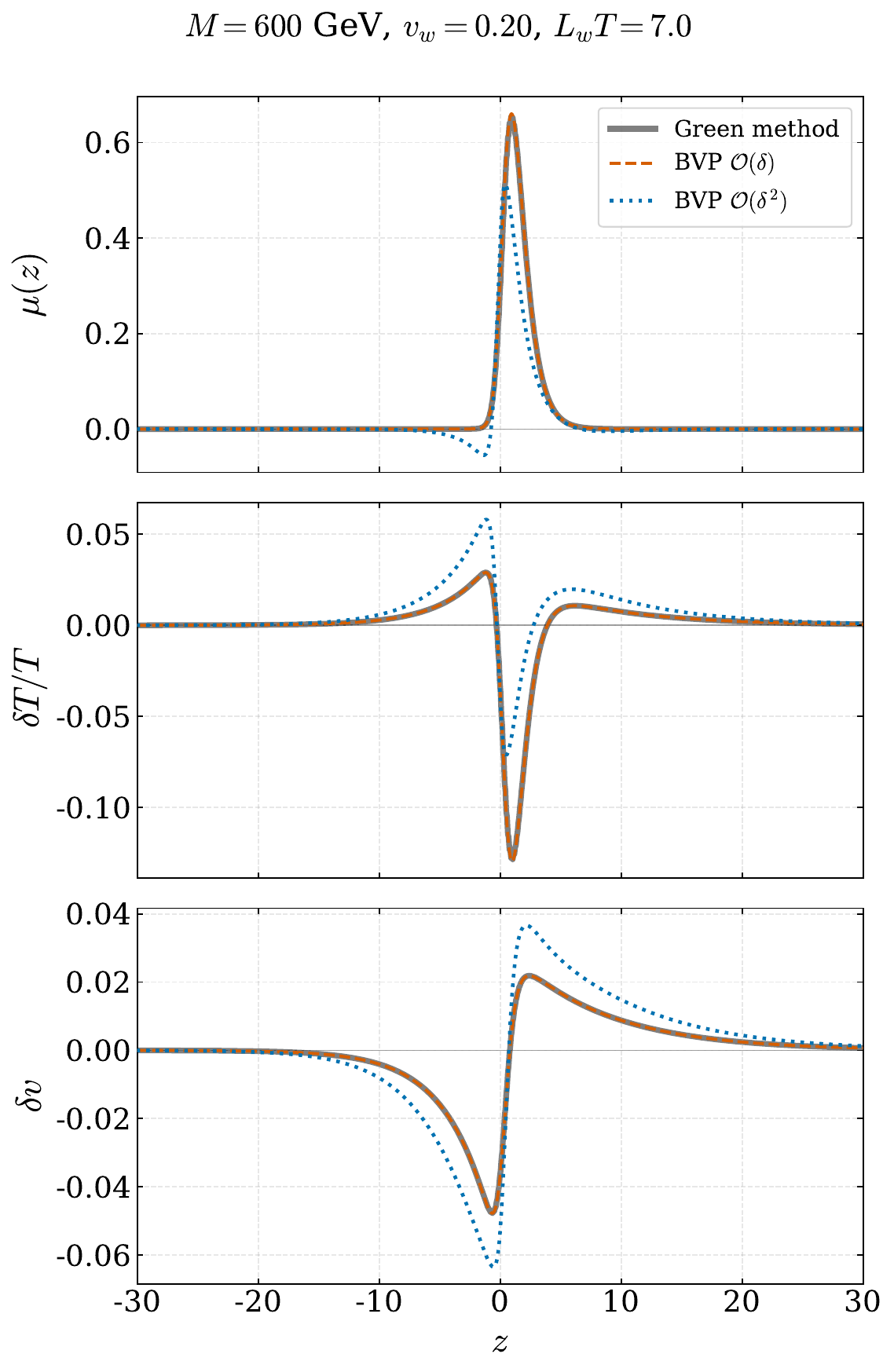}
    \end{subfigure}
    \caption{Spatial profiles of the macroscopic fluid fluctuations $(\mu,\delta T,\delta v)$. The left panel corresponds to a relatively weak phase transition $M=850$ GeV, $v_w=0.2$, where the departures from the equilibrium remain small, and the linear approximation works well. The right panel (b) illustrates a strong phase transition $M= 600$ GeV, $v_w= 0.3$, where larger macroscopic fluctuations have a significant impact when accounting for $\mathcal{O}(\delta ^2)$ contributions. Both plots compare the standard Green's method against the numerical Boundary Value Problem (BVP) solutions at linear $\mathcal{O}(\delta)$ and quadratic $\mathcal{O}(\delta^2)$orders.}
    \label{fig:fluct_second_order}
\end{figure}

Having successfully solved the extended Boltzmann system, we obtain the nonlinear fluid fluctuation profiles $q(z)$ shown in eq.~\eqref{fig:fluct_second_order} that exhibit $\mathcal{O}(\delta^2)$ dynamics.

The next step is to evaluate their backreaction on the scalar field. In the standard linear approach, the departure from equilibrium $\delta f$ inside the out-of-equilibrium integral of eq.~\eqref{eq:KG_equation} is truncated strictly at first order, leading to the friction terms $\mathcal{P}$ and $\mathcal{G}$ as in eq.~\eqref{eq:M1M2}. However, expanding the distribution function as $\delta f =  -f_{eq}' \delta + \frac{1}{2} f_{eq}'' \delta^2 \equiv \delta f_1 +\delta f_2 $ introduces purely second-order corrections to the plasma friction that can be computed as

\begin{equation}
\begin{aligned}
        \mathcal{P} (\delta f_1 + \delta f_2) &= \frac{T^2 N_t}{4 \pi^2 } \int d\xi \frac{d m_t^2}{d\xi} \left( c_{1}\delta\mu + c_{2}\delta T + Y_{\delta^2}(\xi) \right) 
\\       \mathcal{G} (\delta f_1 + \delta f_2) &= \frac{T^2 N_t}{4 \pi^2 } \int d\xi \frac{d m_t^2}{d\xi} \left( c_{1}\delta\mu + c_{2}\delta T + Y_{\delta^2}(\xi) \right) 
        \displaystyle \phi_0\xi     
\end{aligned}
\end{equation}

where $Y_{\delta_2}(\xi)$ is the second order $z$ dependent function that must be integrated:
\begin{equation}
    Y_{\delta^2} (\xi)= \frac{1}{2}\left[ \mu^2(\xi) (-\kappa_1) + (\delta T(\xi))^2 (-\kappa_3) + (\delta v(\xi))^2 \left(-\frac{\kappa_3}{3}\right) + 2\mu(\xi) \delta T(\xi) (-\kappa_2) \right]\, .
\end{equation}
Now one can solve eq.~\eqref{eq:M1M2_full} to find the net impact on $v_w$.

\begin{figure}
    \centering
    %\begin{subfigure}{0.45\textwidth}
    %    \centering
        \includegraphics[width=.45\textwidth]{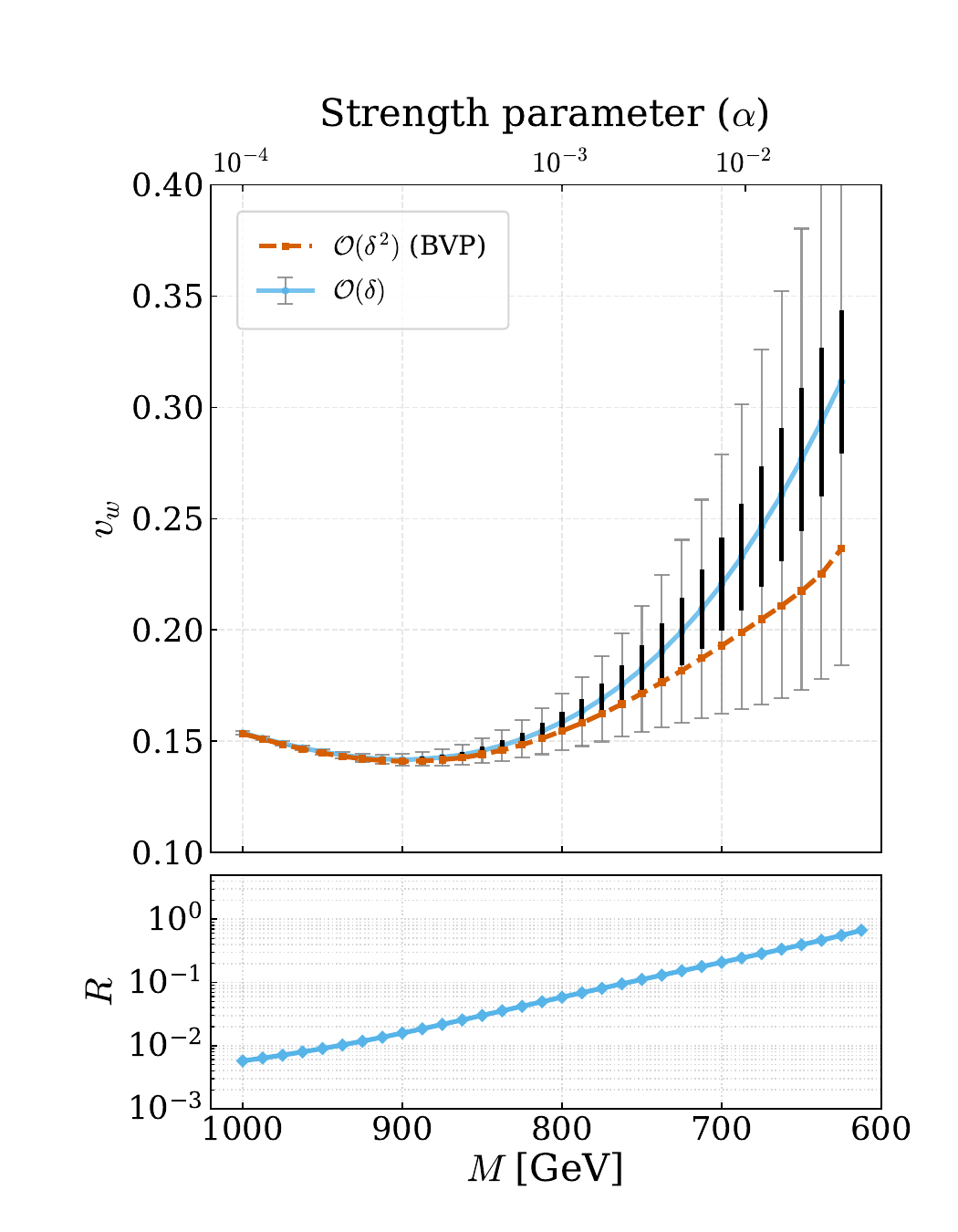}
    %\end{subfigure}
    \qquad
    %\begin{subfigure}{0.45\textwidth}
    %    \centering
        \includegraphics[width=.45\textwidth]{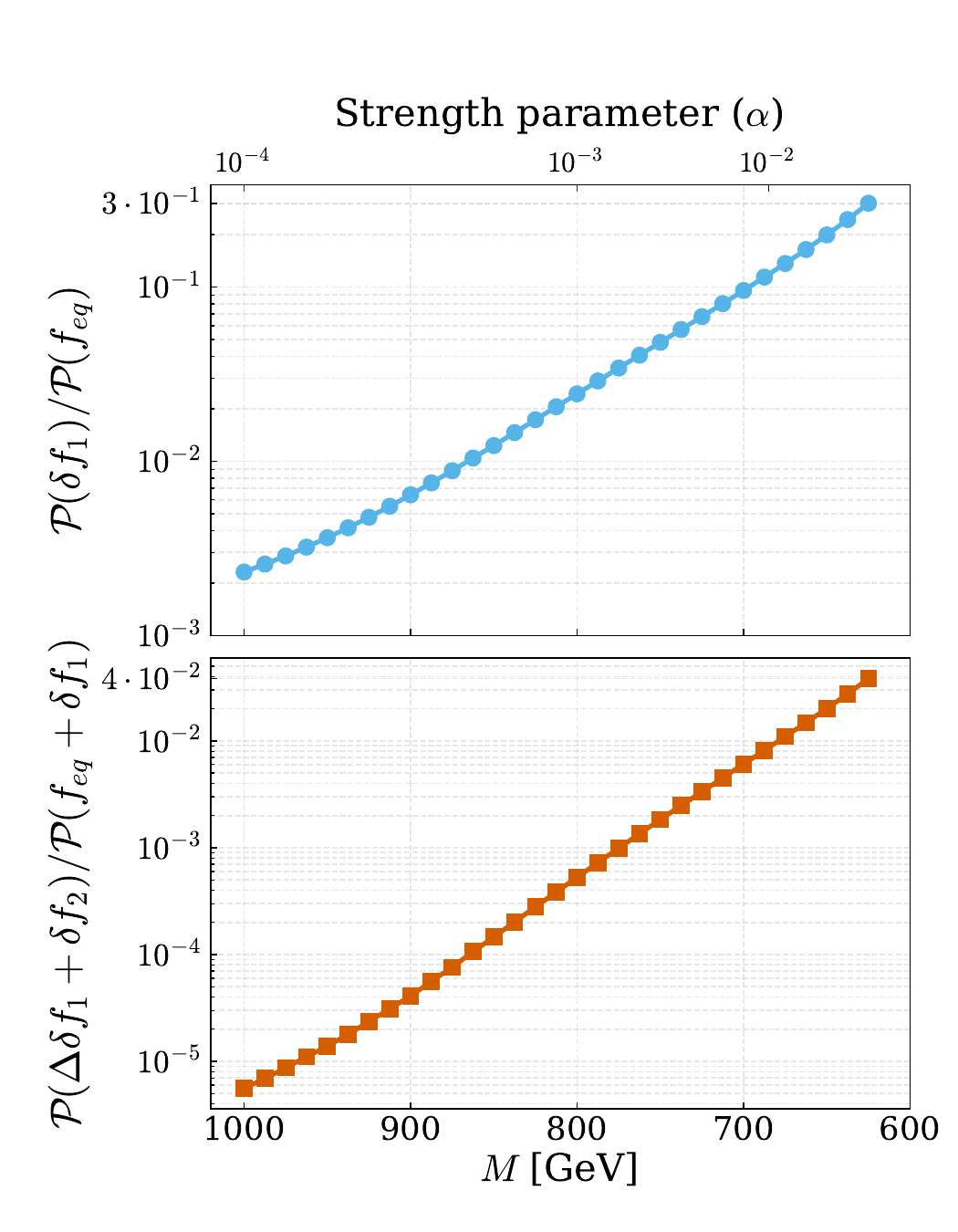}
    %\end{subfigure}
    \caption{Impact of $\mathcal{O}(\delta^2)$ corrections to the bubble wall velocity $v_w$ for the SMEFT model, considering all top quark scattering processes. The left panel compares the standard linear approximation $\mathcal{O}(\delta)$ with the non-linear fluid \emph{Ansatz} solution $\mathcal{O}(\delta^2)$. The dual error bars on the linear solution represent the theoretical uncertainty derived from the Jacobian sensitivity analysis: the thick black inner bars represent the net residual uncertainty (which preserves physical cancellations), while the thinner gray outer bars represent the strictly conservative upper bound (sum of absolute errors). The $\max |R|$ and $\alpha$ associated value with $M$ are also plotted. The right panel displays the pressure ratios associated with the inclusion of $\mathcal{O}(\delta)$ contributions with respect to equilibrium contributions (upper plot), and the lower plot shows the ratio of pressure associated with the inclusion of $\mathcal{O}(\delta^2)$ terms in the equation of motion divided by the $f_{eq}$ and $\delta f$ pressures (solved considering only the $\mathcal{O}(\delta)$ linear system).}
    \label{fig:second_order_vw}
\end{figure}

While solving the extended non-linear system provides the shifted wall velocity $v_w$ and thickness $L_w$ illustrated in the left plot of the figure~\ref{fig:second_order_vw}, one can quantify the uncertainty inherent to the truncation of the fluid \emph{Ansatz}. To establish an error bound on the macroscopic wall parameters, we perform a couple of two-dimensional sensitivity analyses. This method translates the truncation errors of $\delta f_2$ into the deviations $\Delta v_w$ and $\Delta L_w$. To establish an error bound on the macroscopic wall parameters, we decompose the total friction into its linear and quadratic contributions, $\mathcal{P}(\delta f) = \mathcal{P} (\delta f_1) + \mathcal{P}(\delta f_2)$ and $\mathcal{G}(\delta f) = \mathcal{G} (\delta f_1) + \mathcal{G}(\delta f_2)$. This allows us to track the residual forces acting on the moments of eq.~\eqref{eq:M1M2_full} arising from the truncation of the fluid \emph{Ansatz}. We define two distinct metrics to quantify this theoretical uncertainty rigorously:

\paragraph{Net residual uncertainty:} 
This metric evaluates the residual force by preserving the correlations and cancellations between the linear and quadratic fluctuations dictated by the coupled Boltzmann system. It is calculated by taking the absolute value of the net variation in the macroscopic friction,
\begin{equation}
\begin{aligned}
\Delta \mathcal{P}_{\text{net}} &= \left| \mathcal{P}^{(1)}(\delta f_1) - \left[ \mathcal{P}^{(2)}(\delta f_1) + \mathcal{P}^{(2)}(\delta f_2) \right] \right|, \\
\Delta \mathcal{G}_{\text{net}} &= \left| \mathcal{G}^{(1)}(\delta f_1) - \left[ \mathcal{G}^{(2)}(\delta f_1) + \mathcal{G}^{(2)}(\delta f_2) \right] \right|.
\end{aligned}
\label{eq:error_net}
\end{equation}

\paragraph{Strictly conservative upper bound:} 
To account for a worst-case scenario in which all neglected second-order terms constructively add up to push the bubble wall away from equilibrium, we also define a strictly conservative bound. This is obtained by applying the triangle inequality to the individual truncation errors,
\begin{equation}
\begin{aligned}
\Delta \mathcal{P}_{\text{cons}} &= \left| \mathcal{P}^{(1)}(\delta f_1) - \mathcal{P}^{(2)}(\delta f_1) \right| + \left| \mathcal{P}^{(2)}(\delta f_2) \right|, \\
\Delta \mathcal{G}_{\text{cons}} &= \left| \mathcal{G}^{(1)}(\delta f_1) - \mathcal{G}^{(2)}(\delta f_1) \right| + \left| \mathcal{G}^{(2)}(\delta f_2) \right|.
\end{aligned}
\label{eq:error_cons}
\end{equation}
In this conservative formulation, the first term captures the deviation in the first-order friction resulting from the non-linear backreaction of the coupled system, while the second term represents the isolated magnitude of the second-order contributions neglected in the linear approximation. Visually, these bounds translate into dual error bars in our results: $\Delta \mathcal{P}_{\text{net}}$ and $\Delta \mathcal{G}_{\text{net}}$ generate the inner error bars, whereas the strictly conservative constraints $\Delta \mathcal{P}_{\text{cons}}$ and $\Delta \mathcal{G}_{\text{cons}}$ define the wider outer error bands.

To evaluate how the wall velocity changes through these residual forces, we compute the linear response of the moments $M_1$ and $M_2$ of equation~\eqref{eq:M1M2_full}. To analyse that, one can define the Jacobian matrix $J$, calculated numerically via finite differences:
\begin{equation}
    J = \begin{bmatrix}
        \dfrac{\partial M_1}{\partial v_w} & \dfrac{\partial M_1}{\partial L_w} \\[15pt]
        \dfrac{\partial M_2}{\partial v_w} & \dfrac{\partial M_2}{\partial L_w}
    \end{bmatrix}
\end{equation}
The shift required in the wall parameters to restore the steady-state equilibrium $M_1, M_2= 0$ under the influence of the missing forces is then governed by the linear relation
\begin{equation}
J \begin{bmatrix} 
    \Delta v_w \\[6pt] 
    \Delta L_w 
\end{bmatrix} = \begin{bmatrix} 
    - \Delta \mathcal{P} \\[6pt] 
    - \Delta \mathcal{G} 
\end{bmatrix}
\label{eq: Jacobian}
\end{equation}
By inverting the system, we map the residual forces directly to the parameter space $v_w$ and $L_w$.

The macroscopic impact of including second-order $\mathcal{O}(\delta^2)$ non-equilibrium corrections on the terminal bubble wall velocity is summarized in figure~\ref{fig:second_order_vw}.
As predicted for weak phase transitions, the extra friction provided by $\mathcal{O}(\delta^2)$ terms remains minimal. This is a consequence of the small fluctuations in this regime, as illustrated in the left panel of figure~\ref{fig:fluct_second_order}. Consequently, the non-linear fluid \emph{Ansatz} implies negligible modifications to the terminal velocity $v_w$ in this weak regime.

However, the dynamics shift significantly for stronger phase transitions (lower values of $M$ and higher derived transition strength $\alpha$). The right panels of figure~\ref{fig:second_order_vw} provide a detailed pressure ratio analysis. The upper-right plot illustrates that, as the transition becomes stronger, the pressure contribution from the linear out-of-equilibrium terms, $\mathcal{P}(\delta f_1)$, grows to become a substantial fraction of the equilibrium pressure, $\mathcal{P}(f_{eq})$\footnote{$\mathcal{P}(f_{eq})$ here is defined being temperature dependent part of $\Delta V$ in equation~\eqref{eq:deltaV}.}. As pointed-out in ref.~\cite{vandeVis:2025plm}, the condition $\mathcal{P}(\delta f_1)/\mathcal{P}(f_{eq}) \sim \mathcal{O}(1)$ does not necessarily implies the breakdown of the $\delta f_1$ truncation. Therefore, it is crucial to evaluate the impact of the second-order pressure relative to the combined $\mathcal{P}(f_{eq}) + \mathcal{P}(\delta f_1)$. The lower-right plot of figure~\ref{fig:second_order_vw} depicts this purely second-order pressure ratio, $\mathcal{P}(\Delta\delta f_1 + \delta f_2)/(\mathcal{P}(f_{eq})+\mathcal{P}(\delta f_1))$. Even in the strongest regimes evaluated, this ratio remains quantitatively small, reaching a maximum of only around $4 \times 10 ^{-2}$. Here, $\mathcal{P}(\Delta \delta f_1)$ stands for the shift in the $\mathcal{O}(\delta)$ pressure when the $\mathcal{O}(\delta^2)$ terms are dynamically included in the Boltzmann Equation.

Despite the seemingly low magnitude of this additional second-order pressure, its backreaction on the macroscopic wall dynamics is significant. As observed in the left panel of figure~\ref{fig:second_order_vw}, the terminal velocity $v_w$ may significantly deviate from the standard $\mathcal{O}(\delta)$ prediction in the strong transition regime. It is also important to note that this shift goes towards diminishing the wall velocity (i.e. increasing the overall friction), thus \emph{worsening} the agreement with the prediction of \wallgo. 

Crucially, while the true non-linear $\mathcal{O}(\delta^2)$ solution is encompassed by the strictly conservative upper bound (gray outer bars, derived from eq.~\eqref{eq:error_cons}), such large uncertainty bars overestimate the true contribution of the second order (orange dashed line). Furthermore, the non-linear solution falls outside the net error band (thick black bars, derived from eq.~\eqref{eq:error_net}). This indicates that evaluating the net residual forces is insufficient when we approach the strong transition case, as the assumed physical cancellations fail to capture the magnitude of the non-linear backreaction. Together, these observations demonstrate that the linear momentum expansion fundamentally loses its predictive power in the strong phase transition regime. The higher-order terms drastically reshape the steady-state velocity.

We attribute the absence of this pronounced non-linear behavior in the models analyzed by \cite{vandeVis:2025plm} to two main factors. First, the maximum ratio of the $\mathcal{O}(\delta f_2)$ pressure computed by \wallgo in their study was approximately one order of magnitude lower than ours. Secondly, their uncertainty estimate relies on a one-dimensional variation of the pressure with respect to velocity (e.g, $\Delta v_w \approx \Delta \mathcal{P} / (\partial \mathcal{P} / \partial v_w)$), explicitly neglecting the coupled deviations induced by the wall thickness $L_w$ in the Jacobian analysis of eq.~\eqref{eq: Jacobian}. Furthermore, because we are solving a fully coupled system\footnote{\wallgo estimation for $\Delta v_w$ considers the decoupling of the Boltzmann Equation for $\mathcal{O}(\delta)$ and $\mathcal{O}(\delta ^2)$ orders, cf.~\cite{vandeVis:2025plm} for details.}, we directly observe backreaction effects where the $\mathcal{O}(\delta^2)$ dynamics actively modify the first-order  $\delta f_1$ fluctuations, generating therefore the non-negligible $\mathcal{P}(\Delta \delta f_1)$ contribution in the strong transition case.

\section{Conclusions}
\label{sec:conclusions}
In this work we have performed a comprehensive benchmarking of two distinct (and most widespread) numerical frameworks for computing the terminal bubble wall velocity in cosmological phase transitions: the extended fluid \emph{Ansatz} and the spectral method implemented in the \wallgo package. By evaluating the non-equilibrium dynamics within a SMEFT scenario and the Light Higgs Model, we tested and compared the predictions of both approaches across a broad spectrum of transition strengths. Our linear-level analysis revealed a robust agreement between the two methods when including only top-quark annihilation process in the Boltzmann system. For both theoretical frameworks, the terminal velocities computed by the fluid \emph{Ansatz} converged to similar values as the full spectral solutions from \wallgo.

We also showed that this strong correspondence is less remarkable when including multiple interaction channels, or in the regime of strong phase transitions. As the phase transition becomes stronger ($\alpha\gtrsim 10^{-2}$ for the models evaluated here), the extended fluid \emph{Ansatz} develops large macroscopic fluctuations, explicitly signaling a breakdown of the linearization procedure performed in the non-equilibrium distribution function to render the Boltzmann equation into a system of linear differential equations.

 We further investigated this regime by extending the standard construction of the Boltzmann system in the fluid \emph{Ansatz} to incorporate second-order terms (i.e. quadratic in the fluctuations). We found that these higher-order terms may significantly shift the predicted terminal velocity $v_w$, and the more so the stronger the transition becomes. The shift makes the agreement with the prediction of \wallgo even worse than if one would include only the linear terms. Intriguingly, this massive backreaction on the macroscopic wall dynamics occurs even though the direct second-order pressure contributions remains quantitatively minimal compared to the baseline equilibrium and first-order pressures.

Ultimately, these findings expand the analysis regarding the limits of current non-equilibrium modelling techniques for bubble wall velocities. We conclude that while both the extended fluid \emph{Ansatz} and \wallgo serve as computationally efficient tools for weak to moderate phase transitions, the linearized expansion of the fluid \emph{Ansatz} fundamentally breaks down in stronger regimes (as observed in our SMEFT benchmark of $\mathcal{O}(\delta^2)$). However, this mathematical limitation must be contextualized by the fact that, as transitions become stronger, the wall thickness shrinks, compromising the applicability of the fundamental WKB approximation which lies at the basis of the whole Boltzmann approach for computing wall velocities. Furthermore, several theoretical uncertainties discussed in ref.~\cite{vandeVis:2025plm} (which can affect the collision terms by up to an $\mathcal{O}(1)$ factor) currently lack implementation within the fluid \emph{Ansatz} and \wallgo framework. Therefore, achieving robust, high-precision predictions for $v_w$ in strongly transitions will require not only refining the treatment of the sources of uncertainties but, ultimately, moving beyond the semi-classical framework entirely.

\section*{Acknowledgements}

We would like to thank Benoit Laurent for valuable discussions related to the \wallgo \\ project. We also thank Ignacy Na\l\k{e}cz for helpful discussions on the collision terms.
GCD would like to thank CNPq for financial support under grant no. 307565/2025-4.
DAP and ML were supported by the Polish National Science Centre (NCN) grant 2023/50/E/ST2/00177. This work was also supported by the FNP IRA programmes: AstroCeNT (MAB/2018/7), funded from the ERDF, and Astrocent (FENG.02.01-IP.05-A015/25) co-financed by the European Union under FENG 2021–2027; and Teaming for Excellence grant Astrocent Plus (101137080) funded by the European Union with complementary national funding from the MNiSW (MNiSW/2025/DIR/811)..

\appendix
%%%%%%%%%%%%%%%%%%%%%%
\section{Moments from the Boltzmann equation} 
\label{app:sources_en_mom}

In this appendix we present closed-form expressions that allow for the systematic construction of the linearized system of equations at an arbitrary truncation order $\kappa$ of the fluid momentum expansion given in eq.~\eqref{eq:delta}. 

A truncation at order $\kappa$ retains all basis elements weighted by $(p^\mu u_\mu)^{a_i} (p^\nu \bar{u}_\nu)^{b_i}$ subject to the constraint $a_i + b_i \le \kappa$. This procedure yields a total of $D = \frac{(\kappa+1)(\kappa+2)}{2}$ fluctuations, thereby establishing the $D \times D$ dimensionality of the kinetic matrix. 

Since the basis elements are organized into successive polynomial orders $L_i = a_i + b_i$, their enumeration inherently follows a triangular sequence. In order to implement the kinetic matrix and the source terms analytically, we establish a unique mapping between each scalar array index $i \in \{1, \dots, D\}$ and its corresponding physical momentum powers $(a_i, b_i)$ by exploiting the properties of triangular roots:
\begin{equation}
\begin{aligned}
L_i &= \left\lfloor \frac{\sqrt{8i - 7} - 1}{2} \right\rfloor, \\
b_i &= i - \frac{L_i(L_i + 1)}{2} - 1, \\
a_i &= L_i - b_i.
\end{aligned}
\end{equation}

We also define the thermal coefficients $c_n$, $\kappa_n$ which are computed through the integrals over the equilibrium distributions and depend on the quantum statistics of the respective particle species in the plasma. These coefficients are defined generally as
\begin{equation}
 c_n^f\equiv \frac{1}{T^{n+1}} \int_0^\infty p^n (-f^{eq}_f)'\, dp, \qquad \kappa^f_n = \int_0^\infty p^n (-f^{eq}_f)''\, dp,
 \label{eq:c_f}
\end{equation}
where the superscripts $b$ and $f$ stand for bosons and fermions, respectively.

In the massless limit the $c_n$ evaluate analytically to
\begin{equation}
\begin{aligned}
c_n^f &= \left( 1 - \frac{1}{2^{n-1}} \right) n! \zeta_n \qquad   (n\geq 2)
\end{aligned}
\label{eq:constants_c}
\end{equation}
and for $\kappa_n^f$ we have
\begin{equation}
    \kappa_n = - n\, c_{n-1},  \qquad\qquad\ \   (n \ge 2)
\end{equation}
with $\zeta_n$ denoting the Riemann zeta function.

For $n=1$, the inclusion of the particle mass $m$ is required to regulate the infrared behavior, leading to
\begin{equation}
\begin{aligned}
c_1^f &= \ln(2).
\end{aligned}
\end{equation}
The analytical formulas for bosons species $c_n^b$ and $\kappa_n^b$ can be found similarly using the Bose-Einstein distribution function.
\subsection{Kinetic matrix $A$}

To unambiguously determine a generic element $A_{ij}$ of the kinetic matrix, one applies the aforementioned mapping to extract the powers associated with the $i$-th row, $(a_i, b_i)$, and the $j$-th column, $(a_j, b_j)$. Defining the total sum of powers for their intersection as $m = a_i + a_j$ and $n = b_i + b_j$, one evaluates the phase-space integration of the Liouville operator. 

This integration naturally imposes an angular symmetry constraint, such that integrals involving odd powers of the longitudinal momentum strictly vanish. Once the thermal factors are cancelled in the equation, the matrix elements are defined based on the parity of $n$. Absorbing the overall prefactors, the kinetic coefficients can be written in the compact form:
\begin{equation}
A_{ij} = \gamma_w\frac{c_{m+n+2}}{2\pi^2} 
\begin{cases}
\frac{v_w}{n+1}, & \text{if } n \text{ is even,} \\[1ex]
\frac{1}{n+2}, & \text{if } n \text{ is odd,}
\end{cases}
\end{equation}
where $c_{m+n+2}$ correspond to the appropriate thermal coefficients defined above.

\subsection{Source terms $S(\xi)$}

In order to ensure consistency with the dimensionless kinetic equations, the macroscopic source vector $S$ is obtained by evaluating the scaled phase-space source, $\mathcal{S}(p, \xi) \equiv -\mathcal{L}[f^{\text{eq}}]$. By applying the symmetry-reduced directional derivative appropriate for a steady planar wall, one finds
\begin{equation}
\begin{split}
\mathcal{S}(p, \xi) &= - f_{eq}' \gamma_w\Bigg[ v_w T \left( \frac{\partial_\xi m^2}{2T^2} \right) \\
&\quad + \frac{\gamma^2 \partial_\xi v}{T} \Big( v_w (p \cdot u)(p \cdot \bar{u}) + (p \cdot \bar{u})^2 \Big) \\
&\quad - \frac{\partial_\xi T}{T^2} \Big( v_w (p \cdot u)^2 + (p \cdot u)(p \cdot \bar{u}) \Big) \Bigg].
\end{split}
\end{equation}

The $i$-th component of the source vector $S$ is then obtained by projecting this expression onto the momentum basis $(p \cdot u)^{a_i} (p \cdot \bar{u})^{b_i}$, utilizing the index mapping established at the beginning of this appendix. We define the dimensionless integral functional $\mathcal{I}_{m,n}$ as
\begin{equation}
\mathcal{I}_{m,n} \equiv \int \frac{d^3p}{(2\pi)^3 E} \frac{(-f_{eq}')}{T^{m+n+2}} (p \cdot u)^m (p \cdot \bar{u})^n = 
\begin{cases}
\frac{1}{2\pi^2} \frac{c_{m+n+1}}{n+1}, & \text{if } n \text{ is even,} \\
0, & \text{if } n \text{ is odd,}
\end{cases}
\end{equation}
with $c_{m+n+1}$ being the thermal coefficients. Ultimately, the closed-form expression for each dimensionless source component $S_i$ integrated over the phase-space evaluates to
\begin{equation}
\begin{split}
S_i = \gamma_w \Bigg\{ &v_w \left( \frac{\partial_\xi m^2}{2T^2} \right) \mathcal{I}_{a_i, b_i} 
- \frac{\partial_\xi T}{T} \Big[ v_w \mathcal{I}_{a_i+2, b_i} + \mathcal{I}_{a_i+1, b_i+1} \Big] \\
&+ \gamma^2 \partial_\xi v \Big[ v_w \mathcal{I}_{a_i+1, b_i+1} + \mathcal{I}_{a_i, b_i+2} \Big] \Bigg\},
\end{split}
\label{eq:Source}
\end{equation}
where $\partial_\xi T$ and $\partial_\xi v$ denote the spatial derivatives of the plasma temperature and fluid velocity, respectively.

\subsection{Collision terms $\Gamma$}
\label{sec:app_collision}

To systematically evaluate the collision terms, it is highly convenient to define a generic interaction rate valid for any $2 \rightarrow 2$ scattering process at an arbitrary order in the momentum expansion. By linearizing the collision operator and strictly neglecting higher-order $\mathcal{O}(\delta^2)$ perturbations, one can show that the projected collision matrix elements take the form~\cite{Dorsch:2021ubz}
\begin{equation}
\begin{split}
 \mathcal{C}_{pk\rightarrow p'k'} &= \frac{1}{16(2\pi)^8} \int \frac{\, d^3k}{\, E_k} f_{p}^{\text{eq}} f_{k}^{\text{eq}} \\
 &\quad \times \int \frac{d^3p^\prime\, d^3k^\prime}{E_{p^\prime}\, E_{k^\prime}} \bigl(1 + s_{p^\prime} f_{p^\prime}^{\text{eq}}\bigr)\bigl(1 + s_{k^\prime} f_{k^\prime}^{\text{eq}}\bigr) |\mathcal{M}|^2 \delta^{(4)}(p + k - p^\prime - k^\prime) \\
 &\quad \times (p^\mu u_\mu)^{a_i} (p^\nu \bar{u}_\nu)^{b_i} \Big( \delta_p(\xi,p) + \delta_k(\xi,k) - \delta_{p^\prime}(\xi,p^\prime) - \delta_{k^\prime}(\xi,k^\prime) \Big).
\end{split}
\end{equation}

Here, the factor $s_i$ encodes the quantum statistics of the respective particle species, taking $s_i = +1$ for bosons and $s_i = -1$ for fermions. The matrix elements are related to the projections $(p,k,p',k')^{\mu}u_{\mu}$ and $(p,k,p',k')^{\rho}\bar{u}_{\rho}$. To generalize the computation for any process, we isolate these contributions by defining the generalized rate constant after integrating over phase-space:
\begin{equation}
\begin{split}
 \gamma^{pk \to p^\prime k^\prime}_{mnrs} &= \frac{1}{T^{m+n+s+r+4}}\frac{1}{16(2\pi)^8} \int \frac{d^3p\, d^3k}{E_p\, E_k} f_{p}^{\text{eq}} f_{k}^{\text{eq}} \\
 &\quad \times \int \frac{d^3p^\prime\, d^3k^\prime}{E_{p^\prime}\, E_{k^\prime}} \bigl(1 + s_{p^\prime} f_{p^\prime}^{\text{eq}}\bigr)\bigl(1 + s_{k^\prime} f_{k^\prime}^{\text{eq}}\bigr) |\mathcal{M}|^2 \delta^{(4)}(p + k - p^\prime - k^\prime) \\
&\hspace{40pt}\quad \times (p^\mu u_\mu)^{m} (p^\nu \bar{u}_\nu)^{r} \Big( h_p(p^{\mu}u_{\mu})^n(p^{\rho}\bar{u}_{\rho})^s+h_{k}(k^{\mu}u_{\mu})^n(k^{\rho}\bar{u}_{\rho})^s\\
&\hspace{120pt}-h_{p'}(p'^{\mu}u_{\mu})^n(p'^{\rho}\bar{u}_{\rho})^s-h_{k'}(k'^{\mu}u_{\mu})^n(k'^{\rho}\bar{u}_{\rho})^s\Big).
\end{split}
\end{equation}

It is important to emphasize that the linear combination of non-equilibrium fluctuations $\delta$ appearing in the integrand strictly incorporates only the species explicitly participating in the specific Boltzmann equation under consideration (e.g., incoming and outgoing top quarks or $W$ bosons). Consequently, we introduce the projection constant $h_i$, such that $h_i = 1$ for heavy species participating in the Boltzmann equation $(\delta_i \neq 0)$, and $h_i = 0$ for light particles in the diagram leg $i$ $(\delta_i \approx 0)$.

To evaluate the collision integrals efficiently, it is advantageous to adopt a specific spatial parameterization in the center-of-mass (c.o.m.) frame. Let $\theta$ denote the scattering angle between $\vec{p}$ and $\vec{p}\,'$, and $\beta$ the angle between the fluid velocity $\vec{u}$ and the incoming momentum $\vec{p}$. Orienting the coordinate system such that $\hat{z}$ is aligned with $\vec{p}$, $\hat{y}$ is aligned with the component of $\vec{u}$ perpendicular to $\vec{p}$, and $\hat{x}$ is the mutually orthogonal direction, the relevant vectors can be written as
\begin{equation}
\begin{aligned}
\vec{u} &= |\vec{u}| \left( \cos\beta\, \hat{p} + \sin\beta\, \hat{y} \right), \\
\vec{\bar{u}} &= |\vec{\bar{u}}| \left( \cos\alpha\, \hat{p} + \sin\alpha \sin\rho\, \hat{y} + \sin\alpha \cos\rho\, \hat{x} \right), \\
\vec{p}\,' &= |\vec{p}\,'| \left( \cos\theta\, \hat{p} + \sin\theta \sin\phi\, \hat{y} + \sin\theta \cos\phi\, \hat{x} \right), \\
\vec{k} &= |\vec{k}| \left( -\hat{p} \right), \\
\vec{k}\,' &= |\vec{k}\,'| \left( -\cos\theta\, \hat{p} - \sin\theta \sin\phi\, \hat{y} - \sin\theta \cos\phi\, \hat{x} \right).
\end{aligned}
\end{equation}
This kinematic setup allows for a big reduction in the dimension of the of the collision integrals for any $2\rightarrow 2$ process.

The spin-summed scattering amplitude squared is defined as:
$$|\mathcal{M}_{pk \to p'k'}(p, k; p', k')|^2 \equiv \frac{1}{N_p} \sum_{p_i \in p} \sum_{k_i \in k} \sum_{p'_i \in p'} \sum_{k'_i \in k'} |\mathcal{T}_{p_i k_i \to p'_i k'_i}(p, k; p', k')|^2 ,$$
where $\mathcal{T}$ represents the standard transition matrix element. The prefactor $1/N_p$ arises from the definition of $f_p$ as the occupancy of $p$-particles, averaged over the set of internal quantum numbers denoted by $p_i$.

Using the \texttt{FeynCalc} package~\cite{Shtabovenko_2025} we were able to reproduce the amplitudes implemented in the \wallgo code~\cite{Ekstedt:2024fyq}. The only formal distinction is an overall symmetry factor of $1/2$: while \wallgo absorbs this factor into the definition of the collision terms, here it is explicitly included within the amplitude itself. The numerical integration is subsequently performed utilizing the \texttt{Cuba} library~\cite{Hahn_2005}.

Furthermore, our numerical results are in full agreement with those of Laurent and Cline~\cite{Laurent:2020gpg} and also those of ref.~\cite{guiggiani2024bubbledynamicselectroweakscale} in the $\Gamma_{ij}$ entries where there is an expected overlap with the fluid \emph{Ansatz} (upper left $2 \times 2$ block), when care is taken to use the same formulas for the amplitude of the processes in the collision term. It is worth noting that differences in the numerical results of $\Gamma_{ij}$ emerge if we consider distinct analytic formulas at leading-log approximation for $|\mathcal{M}_{pk \to p'k'}|^2$. These differences reflect the inherent theoretical uncertainties of the rough leading-log approximation rather than implementation errors.

To construct the full macroscopic collision matrix $\Gamma$ for the linearized system, we project the collision operator onto the orthogonal momentum basis. Utilizing the index mapping established at the beginning of this appendix, the row index $i$ and column index $j$ uniquely dictate the momentum powers $(a_i, b_i)$ and $(a_j, b_j)$, respectively. The individual elements of the collision matrix $\Gamma_{ij}$ are then directly constructed by evaluating and summing the generic interaction rates over all relevant $2 \rightarrow 2$ scattering processes in the leading-log approximation:
\begin{equation}
\Gamma_{ij} = \sum_{\text{processes}} \gamma^{pk\rightarrow p' k'}_{a_i a_j b_i b_j}.
\label{eq:collision_el}
\end{equation}
To deal with the thermal effects, we use $1/(t-m^2)$ and $1/(u-m^2)$ propagators, where $m$ is the thermal mass of the exchanged particle. This thermal mass effectively regulates the integral, avoiding the otherwise expected divergences in the $u$ and $t$ channels. The thermal mass for the gluon is given by $m_g^2 = 2g_s^2 T^2$ and $m_q^2 = g_s^2T^2/6$ for quarks.
These matrix elements for the top quarks were computed using eq.~\eqref{eq:collision_el} for the following processes: $t\bar{t}\rightarrow gg$ (annihilation of top quarks into gluons), $tq\rightarrow tq$ (scattering with light quarks), and $tg \rightarrow tg $ (scattering with gluons) with the following amplitudes:

\begin{equation}
|\mathcal{M}|^2_{t\bar{t}\rightarrow gg} =  \frac{32 g_s^4}{9}\left[ \frac{t u}{(m_q^2 - t)^2} + \frac{t u}{(m_q^2 - u)^2} \right],
\end{equation}
\begin{equation}
    |\mathcal{M}|^2_{tq\rightarrow tq} = \frac{80}{3}g_s^4 \frac{s^2 + u^2}{(m_g^2 - t)^2},
\end{equation}

\begin{equation}
    |\mathcal{M}|^2_{tg\rightarrow tg} =  16g_s^4 \frac{s^2 + u^2}{(m_g^2 - t)^2} - \frac{64}{9}\frac{su}{(m_q^2-u)^2}.
\end{equation}
The same equivalent matrix elements were implemented in the \wallgo code for a consistent comparison.

\bibliographystyle{JHEP}
\bibliography{LG_refs}

@article{Azatov:2021ifm,
    author = "Azatov, Aleksandr and Vanvlasselaer, Miguel and Yin, Wen",
    title = "{Dark Matter production from relativistic bubble walls}",
    eprint = "2101.05721",
    archivePrefix = "arXiv",
    primaryClass = "hep-ph",
    reportNumber = "SISSA 03/2021/FISI",
    doi = "10.1007/JHEP03(2021)288",
    journal = "JHEP",
    volume = "03",
    pages = "288",
    year = "2021"
}

@article{Baldes:2022oev,
    author = "Baldes, Iason and Gouttenoire, Yann and Sala, Filippo",
    title = "{Hot and heavy dark matter from a weak scale phase transition}",
    eprint = "2207.05096",
    archivePrefix = "arXiv",
    primaryClass = "hep-ph",
    reportNumber = "ULB-TH/22-12",
    doi = "10.21468/SciPostPhys.14.3.033",
    journal = "SciPost Phys.",
    volume = "14",
    pages = "033",
    year = "2023"
}

@article{Giese:2020rtr,
    author = "Giese, Felix and Konstandin, Thomas and van de Vis, Jorinde",
    title = "{Model-independent energy budget of cosmological first-order phase transitions{\textemdash}A sound argument to go beyond the bag model}",
    eprint = "2004.06995",
    archivePrefix = "arXiv",
    primaryClass = "astro-ph.CO",
    reportNumber = "DESY-20-064",
    doi = "10.1088/1475-7516/2020/07/057",
    journal = "JCAP",
    volume = "07",
    number = "07",
    pages = "057",
    year = "2020"
}

@article{Ellis:2018mja,
    author = "Ellis, John and Lewicki, Marek and No, Jos{\'e} Miguel",
    title = "{On the Maximal Strength of a First-Order Electroweak Phase Transition and its Gravitational Wave Signal}",
    eprint = "1809.08242",
    archivePrefix = "arXiv",
    primaryClass = "hep-ph",
    reportNumber = "KCL-PH-TH/2018-46, CERN-TH/2018-197, IFT-UAM/CSIC-18-94, CERN-TH-2018-197",
    doi = "10.1088/1475-7516/2019/04/003",
    journal = "JCAP",
    volume = "04",
    pages = "003",
    year = "2019"
}

@article{Caprini:2019egz,
    author = "Caprini, Chiara and others",
    title = "{Detecting gravitational waves from cosmological phase transitions with LISA: an update}",
    eprint = "1910.13125",
    archivePrefix = "arXiv",
    primaryClass = "astro-ph.CO",
    reportNumber = "DESY-19-159, IPPP/19/27, HIP-2019-14/TH, MITP/19-066, IFT-UAM/CSIC-19-139",
    doi = "10.1088/1475-7516/2020/03/024",
    journal = "JCAP",
    volume = "03",
    pages = "024",
    year = "2020"
}

@article{Giese:2020znk,
    author = "Giese, Felix and Konstandin, Thomas and Schmitz, Kai and van de Vis, Jorinde",
    title = "{Model-independent energy budget for LISA}",
    eprint = "2010.09744",
    archivePrefix = "arXiv",
    primaryClass = "astro-ph.CO",
    reportNumber = "DESY-20-173, DESY 20-173, CERN-TH-2020-170",
    doi = "10.1088/1475-7516/2021/01/072",
    journal = "JCAP",
    volume = "01",
    pages = "072",
    year = "2021"
}

@article{Branchina:2024rva,
    author = "Branchina, Carlo and Conaci, Angela and De Curtis, Stefania and Delle Rose, Luigi and Guiggiani, Andrea and Gil Muyor, Angel and Panico, Giuliano",
    title = "{New calculation of collision integrals for cosmological phase transitions}",
    eprint = "2410.00766",
    archivePrefix = "arXiv",
    primaryClass = "hep-ph",
    doi = "10.1051/epjconf/202431400031",
    journal = "EPJ Web Conf.",
    volume = "314",
    pages = "00031",
    year = "2024"
}

@article{Branchina:2025adj,
    author = "Branchina, Carlo and Conaci, Angela and Delle Rose, Luigi and De Curtis, Stefania",
    title = "{Bubble wall velocity with out-of-equilibrium corrections}",
    eprint = "2510.21942",
    archivePrefix = "arXiv",
    primaryClass = "hep-ph",
    doi = "10.1103/nmkw-7kgk",
    journal = "Phys. Rev. D",
    volume = "113",
    number = "3",
    pages = "035024",
    year = "2026"
}

@article{Caprini:2018mtu,
    author = "Caprini, Chiara and Figueroa, Daniel G.",
    title = "{Cosmological Backgrounds of Gravitational Waves}",
    eprint = "1801.04268",
    archivePrefix = "arXiv",
    primaryClass = "astro-ph.CO",
    doi = "10.1088/1361-6382/aac608",
    journal = "Class. Quant. Grav.",
    volume = "35",
    number = "16",
    pages = "163001",
    year = "2018"
}

@article{Chun:2023ezg,
    author = "Chun, Eung Jin and Dutka, Tomasz P. and Jung, Tae Hyun and Nagels, Xander and Vanvlasselaer, Miguel",
    title = "{Bubble-assisted leptogenesis}",
    eprint = "2305.10759",
    archivePrefix = "arXiv",
    primaryClass = "hep-ph",
    reportNumber = "CTPU-PTC-23-17",
    doi = "10.1007/JHEP09(2023)164",
    journal = "JHEP",
    volume = "09",
    pages = "164",
    year = "2023"
}

@article{Cline:2020jre,
    author = "Cline, James M. and Kainulainen, Kimmo",
    title = "{Electroweak baryogenesis at high bubble wall velocities}",
    eprint = "2001.00568",
    archivePrefix = "arXiv",
    primaryClass = "hep-ph",
    reportNumber = "CERN-TH-2019-227",
    doi = "10.1103/PhysRevD.101.063525",
    journal = "Phys. Rev. D",
    volume = "101",
    number = "6",
    pages = "063525",
    year = "2020"
}

@article{Kainulainen:2024qpm,
    author = "Kainulainen, Kimmo and Venkatesan, Niyati",
    title = "{Systematic moment expansion for electroweak baryogenesis}",
    eprint = "2407.13639",
    archivePrefix = "arXiv",
    primaryClass = "hep-ph",
    doi = "10.1088/1475-7516/2024/08/058",
    journal = "JCAP",
    volume = "08",
    pages = "058",
    year = "2024"
}

@article{DeCurtis:2022hlx,
    author = "De Curtis, Stefania and Rose, Luigi Delle and Guiggiani, Andrea and Muyor, \'Angel Gil and Panico, Giuliano",
    title = "{Bubble wall dynamics at the electroweak phase transition}",
    eprint = "2201.08220",
    archivePrefix = "arXiv",
    primaryClass = "hep-ph",
    doi = "10.1007/JHEP03(2022)163",
    journal = "JHEP",
    volume = "03",
    pages = "163",
    year = "2022"
}

@article{DeCurtis:2023hil,
    author = "De Curtis, Stefania and Delle Rose, Luigi and Guiggiani, Andrea and Gil Muyor, \'Angel and Panico, Giuliano",
    title = "{Collision integrals for cosmological phase transitions}",
    eprint = "2303.05846",
    archivePrefix = "arXiv",
    primaryClass = "hep-ph",
    doi = "10.1007/JHEP05(2023)194",
    journal = "JHEP",
    volume = "05",
    pages = "194",
    year = "2023"
}

@article{DeCurtis:2024hvh,
    author = "De Curtis, Stefania and Delle Rose, Luigi and Guiggiani, Andrea and Gil Muyor, {\'A}ngel and Panico, Giuliano",
    title = "{Non-linearities in cosmological bubble wall dynamics}",
    eprint = "2401.13522",
    archivePrefix = "arXiv",
    primaryClass = "hep-ph",
    doi = "10.1007/JHEP05(2024)009",
    journal = "JHEP",
    volume = "05",
    pages = "009",
    year = "2024"
}

@book{DeGroot:1980dk,
    author = "De Groot, S. R.",
    editor = "Van Leeuwen, W. A. and Van Weert, C. G.",
    title = "{Relativistic Kinetic Theory. Principles and Applications}",
    year = "1980"
}

@article{Dorsch:2021,
    author = "Dorsch, G. C. and Huber, S. J. and Konstandin, T.",
    title = "On the wall velocity dependence of electroweak baryogenesis",
    journal = "JCAP",
    volume = "08",
    pages = "020",
    year = "2021",
    eprint = "2106.06547",
    archivePrefix = "arXiv"
}

@article{Dorsch:2021ubz,
    author = "Dorsch, Glauber C. and Huber, Stephan J. and Konstandin, Thomas",
    title = "{On the wall velocity dependence of electroweak baryogenesis}",
    eprint = "2106.06547",
    archivePrefix = "arXiv",
    primaryClass = "hep-ph",
    reportNumber = "DESY-21-089, DESY 21-089",
    doi = "10.1088/1475-7516/2021/08/020",
    journal = "JCAP",
    volume = "08",
    pages = "020",
    year = "2021"
}

@article{Dorsch:2023tss,
    author = "Dorsch, Glauber C. and Pinto, Daniel A.",
    title = "{Bubble wall velocities with an extended fluid Ansatz}",
    eprint = "2312.02354",
    archivePrefix = "arXiv",
    primaryClass = "hep-ph",
    doi = "10.1088/1475-7516/2024/04/027",
    journal = "JCAP",
    volume = "04",
    pages = "027",
    year = "2024"
}

@article{Dorsch:2024jjl,
    author = "Dorsch, Gl{\'a}uber C. and Konstandin, Thomas and Perboni, Enrico and Pinto, Daniel A.",
    title = "{Non-singular solutions to the Boltzmann equation with a fluid Ansatz}",
    eprint = "2412.09266",
    archivePrefix = "arXiv",
    primaryClass = "hep-ph",
    reportNumber = "DESY-24-193",
    doi = "10.1088/1475-7516/2025/04/033",
    journal = "JCAP",
    volume = "04",
    pages = "033",
    year = "2025"
}

@article{Dorsch_2022,
    title = {A sonic boom in bubble wall friction},
    volume = {2022},
    issn = {1475-7516},
    url = {http://arxiv.org/abs/2112.12548},
    doi = {10.1088/1475-7516/2022/04/010},
    number = {04},
    urldate = {2024-02-09},
    journal = {Journal of Cosmology and Astroparticle Physics},
    author = {Dorsch, Glauber C. and Huber, Stephan J. and Konstandin, Thomas},
    month = apr,
    year = {2022},
    note = {arXiv:2112.12548 [hep-ph]},
    pages = {010},
}

@article{Ekstedt:2024fyq,
    author = "Ekstedt, Andreas and Gould, Oliver and Hirvonen, Joonas and Laurent, Benoit and Niemi, Lauri and Schicho, Philipp and van de Vis, Jorinde",
    title = "{How fast does the WallGo? A package for computing wall velocities in first-order phase transitions}",
    eprint = "2411.04970",
    archivePrefix = "arXiv",
    primaryClass = "hep-ph",
    reportNumber = "CERN-TH-2024-174, DESY-24-162, HIP-2024-21/TH",
    doi = "10.1007/JHEP04(2025)101",
    journal = "JHEP",
    volume = "04",
    pages = "101",
    year = "2025"
}

@article{guiggiani2024bubbledynamicselectroweakscale,
    title={Bubble dynamics at the electroweak scale}, 
    author={Andrea Guiggiani},
    year={2024},
    eprint={2401.18043},
    archivePrefix={arXiv},
    primaryClass={hep-ph},
    url={https://arxiv.org/abs/2401.18043}, 
}

@article{Hahn_2005,
    title={Cuba—a library for multidimensional numerical integration},
    volume={168},
    ISSN={0010-4655},
    url={http://dx.doi.org/10.1016/j.cpc.2005.01.010},
    DOI={10.1016/j.cpc.2005.01.010},
    number={2},
    journal={Computer Physics Communications},
    publisher={Elsevier BV},
    author={Hahn, T.},
    year={2005},
    month=jun, pages={78–95} 
}

@inproceedings{Heffernan:2026kva,
    author = "Heffernan, Anna and others",
    title = "{LISA and the LISA Science Team}",
    eprint = "2601.15365",
    archivePrefix = "arXiv",
    primaryClass = "astro-ph.IM",
    month = "1",
    year = "2026"
}

@article{Jiang:2023nkj,
    author = "Jiang, Siyu and Huang, Fa Peng and Li, Chong Sheng",
    title = "{Hydrodynamic effects on the filtered dark matter produced by a first-order phase transition}",
    eprint = "2305.02218",
    archivePrefix = "arXiv",
    primaryClass = "hep-ph",
    doi = "10.1103/PhysRevD.108.063508",
    journal = "Phys. Rev. D",
    volume = "108",
    number = "6",
    pages = "063508",
    year = "2023"
}

@article{Konstandin:2013caa,
    author = "Konstandin, Thomas",
    title = "{Quantum Transport and Electroweak Baryogenesis}",
    eprint = "1302.6713",
    archivePrefix = "arXiv",
    primaryClass = "hep-ph",
    reportNumber = "DESY-13-036",
    doi = "10.3367/UFNe.0183.201308a.0785",
    journal = "Phys. Usp.",
    volume = "56",
    pages = "747--771",
    year = "2013"
}

@article{Konstandin_2014,
    title={From Boltzmann equations to steady wall velocities},
    volume={2014},
    ISSN={1475-7516},
    url={http://dx.doi.org/10.1088/1475-7516/2014/09/028},
    DOI={10.1088/1475-7516/2014/09/028},
    number={09},
    journal={Journal of Cosmology and Astroparticle Physics},
    publisher={IOP Publishing},
    author={Konstandin, Thomas and Nardini, Germano and Rues, Ingo},
    year={2014},
    month=sep, pages={028–028} 
}

@article{Laurent:2020gpg,
    author = "Laurent, Benoit and Cline, James M.",
    title = "{Fluid equations for fast-moving electroweak bubble walls}",
    eprint = "2007.10935",
    archivePrefix = "arXiv",
    primaryClass = "hep-ph",
    doi = "10.1103/PhysRevD.102.063516",
    journal = "Phys. Rev. D",
    volume = "102",
    number = "6",
    pages = "063516",
    year = "2020"
}

@article{Laurent:2022jrs,
    author = "Laurent, Benoit and Cline, James M.",
    title = "{First principles determination of bubble wall velocity}",
    eprint = "2204.13120",
    archivePrefix = "arXiv",
    primaryClass = "hep-ph",
    doi = "10.1103/PhysRevD.106.023501",
    journal = "Phys. Rev. D",
    volume = "106",
    number = "2",
    pages = "023501",
    year = "2022"
}

@article{Lewicki:2021pgr,
    author = "Lewicki, Marek and Merchand, Marco and Zych, Mateusz",
    title = "{Electroweak bubble wall expansion: gravitational waves and baryogenesis in Standard Model-like thermal plasma}",
    eprint = "2111.02393",
    archivePrefix = "arXiv",
    primaryClass = "astro-ph.CO",
    doi = "10.1007/JHEP02(2022)017",
    journal = "JHEP",
    volume = "02",
    pages = "017",
    year = "2022"
}

@article{LIGOScientific:2016aoc,
    author = "Abbott, B. P. and others",
    collaboration = "LIGO Scientific, Virgo",
    title = "{Observation of Gravitational Waves from a Binary Black Hole Merger}",
    eprint = "1602.03837",
    archivePrefix = "arXiv",
    primaryClass = "gr-qc",
    reportNumber = "LIGO-P150914",
    doi = "10.1103/PhysRevLett.116.061102",
    journal = "Phys. Rev. Lett.",
    volume = "116",
    number = "6",
    pages = "061102",
    year = "2016"
}

@article{LIGOScientific:2018mvr,
    author = "Abbott, B. P. and others",
    collaboration = "LIGO Scientific, Virgo",
    title = "{GWTC-1: A Gravitational-Wave Transient Catalog of Compact Binary Mergers Observed by LIGO and Virgo during the First and Second Observing Runs}",
    eprint = "1811.12907",
    archivePrefix = "arXiv",
    primaryClass = "astro-ph.HE",
    reportNumber = "LIGO-P1800307",
    doi = "10.1103/PhysRevX.9.031040",
    journal = "Phys. Rev. X",
    volume = "9",
    number = "3",
    pages = "031040",
    year = "2019"
}

@article{LIGOScientific:2021djp,
    author = "Abbott, R. and others",
    collaboration = "LIGO Scientific, VIRGO, KAGRA",
    title = "{GWTC-3: Compact Binary Coalescences Observed by LIGO and Virgo During the Second Part of the Third Observing Run}",
    eprint = "2111.03606",
    archivePrefix = "arXiv",
    primaryClass = "gr-qc",
    reportNumber = "LIGO-P2000318",
    month = "11",
    year = "2021"
}

@article{LIGOScientific:2021usb,
    author = "Abbott, R. and others",
    collaboration = "LIGO Scientific, VIRGO",
    title = "{GWTC-2.1: Deep Extended Catalog of Compact Binary Coalescences Observed by LIGO and Virgo During the First Half of the Third Observing Run}",
    eprint = "2108.01045",
    archivePrefix = "arXiv",
    primaryClass = "gr-qc",
    reportNumber = "LIGO-P2100063",
    month = "8",
    year = "2021"
}

@article{LISA:2017pwj,
    author = "Amaro-Seoane, Pau and others",
    collaboration = "LISA",
    title = "{Laser Interferometer Space Antenna}",
    eprint = "1702.00786",
    archivePrefix = "arXiv",
    primaryClass = "astro-ph.IM",
    month = "2",
    year = "2017"
}

@article{Moore_1995,
    title={How fast can the wall move? A study of the electroweak phase transition dynamics},
    volume={52},
    ISSN={0556-2821},
    url={http://dx.doi.org/10.1103/PhysRevD.52.7182},
    DOI={10.1103/physrevd.52.7182},
    number={12},
    journal={Physical Review D},
    publisher={American Physical Society (APS)},
    author={Moore, Guy D. and Prokopec, Tomislav},
    year={1995},
    month=Dec, pages={7182–7204} 
}

@article{Fromme:2006wx,
    author = "Fromme, Lars and Huber, Stephan J.",
    title = "{Top transport in electroweak baryogenesis}",
    eprint = "hep-ph/0604159",
    archivePrefix = "arXiv",
    reportNumber = "CERN-PH-TH-2006-064, BI-TP-2006-10",
    doi = "10.1088/1126-6708/2007/03/049",
    journal = "JHEP",
    volume = "03",
    pages = "049",
    year = "2007"
}

@article{Arnold:2000dr,
    author = "Arnold, Peter Brockway and Moore, Guy D. and Yaffe, Laurence G.",
    title = "{Transport coefficients in high temperature gauge theories. 1. Leading log results}",
    eprint = "hep-ph/0010177",
    archivePrefix = "arXiv",
    reportNumber = "UW-PT-00-15",
    doi = "10.1088/1126-6708/2000/11/001",
    journal = "JHEP",
    volume = "11",
    pages = "001",
    year = "2000"
}

@article{Morrissey:2012db,
    author = "Morrissey, David E. and Ramsey-Musolf, Michael J.",
    title = "{Electroweak baryogenesis}",
    eprint = "1206.2942",
    archivePrefix = "arXiv",
    primaryClass = "hep-ph",
    reportNumber = "NPAC-12-08",
    doi = "10.1088/1367-2630/14/12/125003",
    journal = "New J. Phys.",
    volume = "14",
    pages = "125003",
    year = "2012"
}

@article{NANOGrav:2023gor,
    author = "Agazie, Gabriella and others",
    collaboration = "NANOGrav",
    title = "{The NANOGrav 15 yr Data Set: Evidence for a Gravitational-wave Background}",
    eprint = "2306.16213",
    archivePrefix = "arXiv",
    primaryClass = "astro-ph.HE",
    doi = "10.3847/2041-8213/acdac6",
    journal = "Astrophys. J. Lett.",
    volume = "951",
    number = "1",
    pages = "L8",
    year = "2023"
}

@article{Shtabovenko_2025,
    title={FeynCalc 10: Do multiloop integrals dream of computer codes?},
    volume={306},
    ISSN={0010-4655},
    url={http://dx.doi.org/10.1016/j.cpc.2024.109357},
    DOI={10.1016/j.cpc.2024.109357},
    journal={Computer Physics Communications},
    publisher={Elsevier BV},
    author={Shtabovenko, Vladyslav and Mertig, Rolf and Orellana, Frederik},
    year={2025},
    month=jan, pages={109357} 
}

@article{vandeVis:2025plm,
    author = "van de Vis, Jorinde and Schicho, Philipp and Niemi, Lauri and Laurent, Benoit and Hirvonen, Joonas and Gould, Oliver",
    title = "{WallGo investigates: Theoretical uncertainties in the bubble wall velocity}",
    eprint = "2510.27691",
    archivePrefix = "arXiv",
    primaryClass = "hep-ph",
    reportNumber = "CERN-TH-2025-221",
    doi = "10.1007/JHEP04(2026)041",
    journal = "JHEP",
    volume = "04",
    pages = "041",
    year = "2026"
}
\end{document}